\documentclass[sigconf,screen	]{acmart}

\AtBeginDocument{%
  \providecommand\BibTeX{{%
    \normalfont B\kern-0.5em{\scshape i\kern-0.25em b}\kern-0.8em\TeX}}}

\setcopyright{acmcopyright}
\copyrightyear{2020}
\acmYear{2020}
\acmConference[ICSE '20]{42nd International Conference on Software Engineering}{May 23--29, 2020}{Seoul, Republic of Korea}
\acmBooktitle{42nd International Conference on Software Engineering (ICSE '20), May 23--29, 2020, Seoul, Republic of Korea}
\acmPrice{15.00}
\acmDOI{10.1145/3377811.3380338}
\acmISBN{978-1-4503-7121-6/20/05}

\usepackage[ruled,vlined,lined,commentsnumbered]{algorithm2e}
\usepackage{booktabs}
\usepackage{moreverb}
\usepackage{fontenc}
\usepackage{amsmath}
\usepackage{amssymb}
\usepackage{fancybox}
\usepackage{color}
\usepackage{colortbl}
\usepackage{array}
\usepackage{multirow}
\usepackage{multicol}
\usepackage{listings}
\usepackage{makecell}
\usepackage{graphicx}
\usepackage{setspace}
\usepackage{soul}

\usepackage{balance}
\usepackage{csvsimple}
\usepackage{longtable}
\usepackage[skip=0pt,font=small,labelfont=bf]{caption}
\usepackage{url}
\usepackage{lscape}
\usepackage{rotating}
\usepackage{tikz}
\usepackage{enumitem}
\usepackage{subcaption}

\usepackage[most]{tcolorbox}

\usepackage{threeparttable}

\usepackage{marvosym}
\usepackage{pifont}

\newcolumntype{L}[1]{>{\raggedright\let\newline\\\arraybackslash\hspace{0pt}}m{#1}}
\newcolumntype{C}[1]{>{\centering\let\newline\\\arraybackslash\hspace{0pt}}m{#1}}
\newcolumntype{R}[1]{>{\raggedleft\let\newline\\\arraybackslash\hspace{0pt}}m{#1}}

\definecolor{codegreen}{rgb}{0,0.6,0}
\definecolor{codegray}{rgb}{0.5,0.5,0.5}
\definecolor{codepurple}{rgb}{0.58,0,0.82}
\definecolor{backcolour}{rgb}{0.95,0.95,0.92}

\lstdefinestyle{mystyle}{
    commentstyle=\color{codegreen},
    keywordstyle=\color{magenta},
    numberstyle=\tiny\color{codegray},
    stringstyle=\color{codepurple},
    basicstyle=\footnotesize,
    breakatwhitespace=false,
    breaklines=true,
    captionpos=b,
    keepspaces=true,
    showspaces=false,
    showstringspaces=false,
    showtabs=false,
    tabsize=2
}

\setlist{noitemsep} 

\definecolor{darkpastelred}{rgb}{0.76, 0.23, 0.13}
\definecolor{ao(english)}{rgb}{0.0, 0.5, 0.0}

\definecolor{darkpastelred}{rgb}{0.76, 0.23, 0.13}
\definecolor{ao(english)}{rgb}{0.0, 0.5, 0.0}

\lstdefinelanguage{diff}{
  morecomment=[f][\color{blue}]{@@},     
  morecomment=[f][\color{red}]-,         
  morecomment=[f][\color{codegreen}]+,       
  morecomment=[f][\color{red}]{---}, 
  morecomment=[f][\color{codegreen}]{+++},
}

\definecolor{yellow}{RGB}{255,255,153}
\definecolor{grey}{RGB}{224,224,224}

\newboolean{showcomments}
\setboolean{showcomments}{true}
\ifthenelse{\boolean{showcomments}}
 { \newcommand{\mynote}[2]{
      \fbox{\bfseries\sffamily\scriptsize#1}
        {\small$\blacktriangleright$\textsf{\emph{#2}}$\blacktriangleleft$}}}
        { \newcommand{\mynote}[2]{}}

\definecolor{DarkOrange}{rgb}{0.8,0.3,0.0}
\definecolor{DarkCyan}{rgb}{0.0, 0.55, 0.55}

\newcolumntype{?}{!{\vrule width 1pt}}

\newcommand{\nbugs}{395\xspace}

\newcommand{\notez}[1]{
\begin{tcolorbox}[tile,size=fbox,boxsep=1.1mm,boxrule=0pt,top=0pt,bottom=0pt,
borderline west={0.8mm}{0pt}{black!50!white},colback=black!5!white]
\em #1
\end{tcolorbox}
}

\begin{document}

\title[On the Efficiency of Test Suite based Program Repair\\A Systematic Assessment of 16 Automated Repair Systems for Java Programs]{
On the Efficiency of Test Suite based Program Repair
}
\subtitle{A Systematic Assessment of 16 Automated Repair Systems for Java Programs}

\settopmatter{printacmref=true}

\author{Kui Liu}
\email{brucekuiliu@gmail.com}
\additionalaffiliation{%
   \institution{University of Luxembourg}
}
\affiliation{%
   \institution{Nanjing University of Aeronautics and Astronautics}
 	\country{China}
}

\author{Shangwen Wang}
\authornote{Co-first author and corresponding author.}
\email{wangshangwen13@nudt.edu.cn}
\affiliation{%
   \institution{National University of Defense Technology}
 	\country{China}
}

\author{Anil Koyuncu}
\author{Kisub Kim}
\email{{anil.koyuncu, kisub.kim}@uni.lu}
\affiliation{%
   \institution{University of Luxembourg}
 	\country{Luxembourg}
}

\author{Tegawend{\'e} F. Bissyand{\'e}}
\email{tegawende.bissyande@uni.lu}
\affiliation{%
   \institution{University of Luxembourg}
 	\country{Luxembourg}
}

\author{Dongsun Kim}
\email{darkrsw@furiosa.ai}
\affiliation{%
   \institution{Furiosa.ai}
   \country{Republic of Korea}
}

\author{Peng Wu}
\email{wupeng15@nudt.edu.cn}
\affiliation{%
   \institution{National University of Defense Technology}
 	\country{China}
}

\author{Jacques Klein}
\email{jacques.klein@uni.lu}
\affiliation{%
   \institution{University of Luxembourg}
 	\country{Luxembourg}
}

\author{Xiaoguang Mao}
\email{xgmao@nudt.edu.cn}
\affiliation{%
   \institution{National University of Defense Technology}
 	\country{China}
}

\author{Yves Le Traon}
\email{yves.letraon@uni.lu}
\affiliation{%
   \institution{University of Luxembourg}
   \country{Luxembourg}
}

\renewcommand{\shortauthors}{Kui Liu, Shangwen Wang, Anil Koyuncu, Kisub Kim, Tegawendé F. Bissyandé,\\ Dongsun Kim, Peng Wu, Jacques Klein, Xiaoguang Mao, and Yves Le Traon}

\begin{abstract} Test-based automated program repair has been a prolific field
of research in software engineering in the last decade. Many approaches have
indeed been proposed, which leverage test suites as a weak, but affordable,
approximation to program specifications. Although the literature regularly sets
new records on the number of benchmark bugs that can be fixed, several studies
increasingly raise concerns about the limitations and biases of state-of-the-art
approaches. For example, the {\em correctness} of generated patches has been
questioned in a number of studies, while other researchers pointed out that
evaluation schemes may be misleading with respect to the processing of fault
localization results. Nevertheless, there is little work addressing the
efficiency of patch generation, with regard to the practicality of program repair.
In this paper, we fill this gap in the literature, by providing an extensive
review on the efficiency of test suite based program repair. Our objective is to assess
the number of generated patch candidates, since this information is correlated
to (1) the strategy to traverse the search space efficiently in order to select
{\em sensical} repair attempts, (2) the strategy to minimize the test effort for
identifying a {\em plausible} patch, (3) as well as the strategy to prioritize
the generation of a {\em correct} patch. To that end, we perform a large-scale
empirical study on the efficiency, in terms of quantity of generated patch
candidates of the 16 open-source repair tools for Java programs. The
experiments are carefully conducted under the same fault localization configurations to limit biases.
Eventually, among other findings, we note that: (1) many irrelevant patch
candidates are generated by changing wrong code locations; (2)
however, if the search space is carefully triaged, fault localization noise has
little impact on patch generation efficiency; (3) yet, current template-based
repair systems, which are known to be most effective in fixing a large number
of bugs, are actually least efficient as they tend to generate majoritarily
irrelevant patch candidates.
\end{abstract}



\maketitle


\vspace{-0.9mm}
\section{Introduction}
\label{sec:intro}

\begin{figure}[!h]
	\centering
    \includegraphics[width=\linewidth]{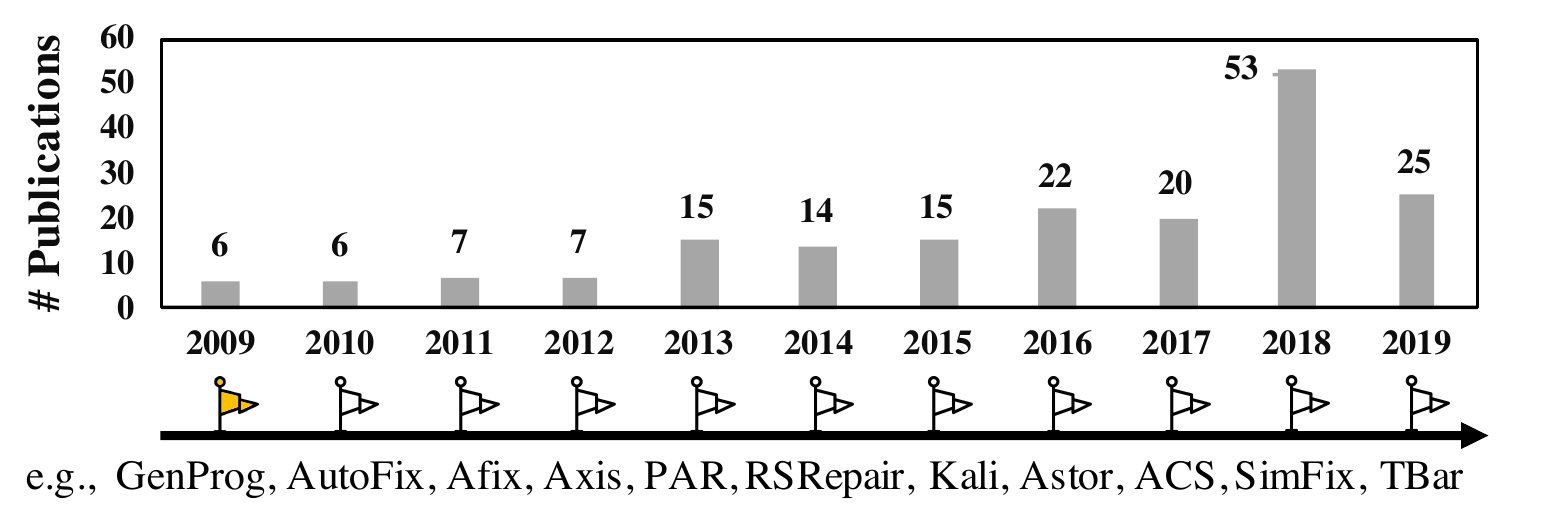}
    \vspace{-4mm}
    \caption{APR research publications since \protect2009\protect\footnotemark.}
    \label{fig:10years}
\end{figure}

In the last decade, Automated Program Repair (APR)~\cite{gazzola2019automatic,
monperrus2018automatic, le2019automated} has extensively grown as a prominent
research topic in the software engineering community. Figure~\ref{fig:10years}
overviews the research activities of this topic. The associated literature
includes a broad range of techniques that use heuristics (e.g., via random
mutation operations~\cite{le2012genprog}), constraints solving (e.g., via symbolic execution~\cite{nguyen2013semfix}), or
machine learning (e.g., via building a code transformation model~\cite{gupta2017deepfix}) to drive
patch generation. A living review of automated program repair research appears
in~\cite{monperrus2018living}, which shows that the research in this field has
been revived with the seminal work, ten years ago, of Weimer et
al.~\cite{weimer2009automatically} on {\em generate-and-validate} approaches.
Patches are generated to be applied on a buggy program until the patched program
meets the desired behaviour. In the absence of formal specifications of the desired
behaviour, test suites are leveraged as  {\em affordable partial specifications}
for validating generated patches. Over the years, the community has
incrementally advanced the state-of-the-art with numerous test-based approaches
that have been shown effective in generating valid patches for a significant
fraction of defects within well-established
benchmarks~\cite{just2014defects4j,madeiral2019bears,lin2017quixbugs,saha2018bugs}.

~\footnotetext{Data are extracted from Monperrus's living review on APR~\cite{monperrus2018living}.}

\vspace{-5mm}
Several studies have revisited the constraints and performance of program
repair systems, and have thus contributed to shaping research directions towards
improving the state-of-the-art. For example, Qi et al.~\cite{qi2015analysis}
have early shown that repair tools generate mostly overfitting patches (i.e.,
patches that pass the incomplete test suites) but are actually incorrect. Their
study led to assessment results being now carefully presented in a way that
highlights the capability of new approaches to correctly repair programs.
Motwani et al.~\cite{motwani2018automated} then questioned whether
state-of-the-art approaches can deal with hard and important bugs. Liu
et al.~\cite{liu2019you}  recently revealed significant biases with fault
localization configurations in APR system evaluations. More recently, Durieux
et al.~\cite{durieux2019empirical} have shown that state-of-the-art tools may
actually be overfitting the associated study benchmarks.

Performance measurement of repair systems has evolved to progressively consider the
number of correctly-fixed bugs or the diversity of benchmark bugs~\cite{durieux2019empirical} that are
fixed. Another performance aspect that deserves investigation is the {\em
efficiency} of the patch generation system.  It is however mentioned in only a
few assessment reports~\cite{xiong2017precise, ghanbari2019practical}. Yet, efficiency is a key property
for bringing program repair into general use within practitioners' settings.
Indeed, APR aims to alleviate the manual effort involved in resolving software
bugs, and holds this promise in two scenarios: in production, it is expected to
drastically reduce the time-to-fix delays and minimize downtime; in a
development cycle, APR can help suggest changes to accelerate debugging. Yet,
until now, literature
approaches~\cite{jiang2018shaping,liu2019tbar,saha2019harnessing,liu2019tbar,xiong2017precise}
have mainly focused on highlighting the increased performance on {\em eventually}
fixing more and more benchmark bugs. In recent work, Ghanbari et
al.~\cite{ghanbari2019practical} raised the efficiency issue and built on the
time cost criterion to demonstrate the efficiency of their PraPR tool (which
does not require re-compiling source code). This criterion, which was already
mentioned in a few of the previous work~\cite{xiong2017precise,
wen2018context,liu2018mining}, however, has limitations with respect to
generalizability (cf. Section~\ref{sec:bg}): execution time is (1) dependent on
many variables that are unrelated to the approach implemented in the repair
system; and (2) is generally unstable.

We postulate that the efficiency of
test-based program repair should be assessed along with the following question: {\bf
how many attempts does the repair system make before catching a valid patch?} In previous work, Qi et al.~\cite{qi2013using} have formulated this
question into a metric that served to assess the effectiveness of fault
localization techniques in a platform-agnostic manner. To the best of our
knowledge, little attention has been paid to measuring repair efficiency by
estimating the number of validated patch candidates.

In this paper, we report on the results of a large scale empirical study on the efficiency of
test-based program repair systems. Our study considers 16 APR systems targeting
Java programs, and performs a systematic assessment under identical
and controlled fault localization configurations. The objective of this work is
to contribute a comprehensive analysis of repair efficiency to the literature
with respect to generated patches for a large spectrum of APR systems. Eventually,
we gather insights on how the strategies of approaches in the literature
affect repair efficiency. Overall, we mainly find that:
\begin{itemize}
	\item [F0:] So far, efficiency is not a widely-valued performance target. We found that state-of-the-art APR tools are the least efficient. This calls for an industry investigation of the impact of efficiency on adoption (or lack thereof).
	\item [F1:] Across time, repair tools subsume each other in terms of which benchmark bugs can be fixed. Unfortunately, effectiveness (i.e., how many bugs are eventually fixed) is increased at the expense of efficiency (i.e., how many repair attempts are made before a given bug is fixed).
	\item [F2:] Template-based repair systems are generally inefficient as they produce too many patch candidates. However, when the templates are mined from clean datasets or are specialized to specific bugs, efficiency can be substantially improved.
	\item [F3:] Literature approaches develop a few strategies, such as constraint solving or donor code search, which contribute to drastically reducing the nonsensical or in-plausible patches. 
	\item [F4:] APR systems that implement random search over the repair search space require large sets of patch candidates to increase the likelihood of hitting a correct patch.
	\item [F5:] Implementation details can diversely influence the repair efficiency of an APR approach.
\end{itemize}

\section{Background and Motivation}
\label{sec:bg}

\begin{figure}[!tp]
    \centering
    \includegraphics[width=\linewidth]{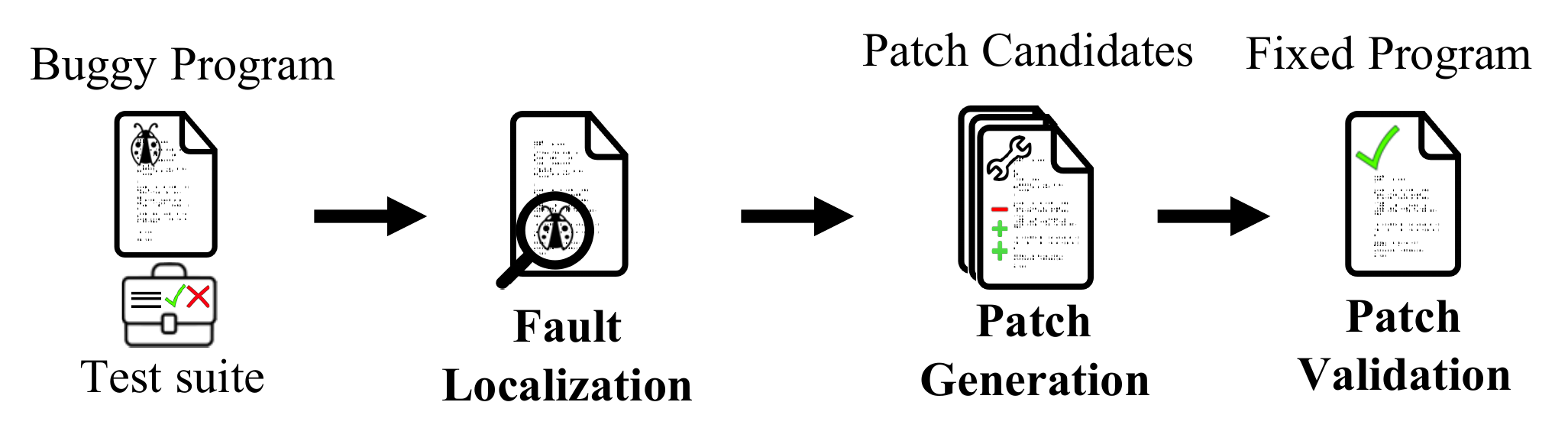}
    \vspace{-4mm}
    \caption{Standard steps in a pipeline of Automated Program Repair.}
    \label{fig:APRPro}
\end{figure}

Test suite based program repair systems commonly implement a three-step pipeline as illustrated in Figure~\ref{fig:APRPro}: {\bf fault localization},
which produces a ranked list of suspicious code locations that should be modified to fix the bug;
{\bf patch generation}, which implements the change operators that are applied on the buggy code locations;
and {\bf patch validation}, which executes the test cases to check
that the patched program meets the behaviour (approximatively) specified by the test suite.

If a patch candidate can pass all the given test cases (both previously-passing and
previously-failing test cases on the buggy version), it is regarded as a {\bf valid} patch.
This criterion was first used by Weimer et al.~\cite{weimer2009automatically}
in their seminal work on GenProg, and has become the de-facto metric of repair performance~\cite{le2019automated}.
Nevertheless, as later studies have revealed,
even if a generated patch can pass all test cases, it might break a necessary
behaviour or introduce other faults, which are not covered by the given test suite~\cite{smith2015cure}.
Besides, a developer may not accept the patch due to several reasons
such as coding convention~\cite{kim2013automatic,monperrus2014critical}.
All such valid patches in terms of the test suite are therefore now referred to as {\bf plausible} since they require further
investigations to ensure that they are {\bf correct}, i.e., acceptable to developers.
In the literature, {\em correctness} is generally assessed manually by comparing the APR-generated patch against the developer-provided patch available in the benchmark.

\notez{Studies in the literature, such as the recent work of Durieux et al.~\cite{durieux2019empirical} on benchmark overfitting, generally focus on information about plausible patches given that correctness is hard to assess.
Our work is the first to explore artifacts from the literature, where researchers provide correctness labels of their generated patches, in order to extract and categorize implicit rules used by the community to define correctness.
We expect that these rules will be studied and augmented by the community to enable systematic assessment of correctness.}

Efficiency of APR tools has been assessed in the literature~\cite{xiong2017precise,wen2018context,hua2018towards,ghanbari2019practical} via measuring the time-to-generate-and-validate patches. Table~\ref{tab:PraPRTimeCost} presents the time cost of the {\tt PraPR}~\cite{ghanbari2019practical}  state-of-the-art repair tool on Defects4J~\cite{just2014defects4j} program samples.
On average, for each {\tt Closure} bug, {\tt PraPR} generated and validated more than 29 thousand patches, approximately 10 times more than the average number of patches that are generated and validated for each {\tt Chart} bug.
Yet, the time cost for {\tt Closure} bugs is ~20 times more than the time cost for {\tt Chart} bugs. This suggests that it is challenging to define a generically-suitable time budget for repairing bugs. We further note that correlation tests did not reveal any linear correlation between the time cost of repairing a bug and benchmark properties such as the number of test cases or program sizes. Consequently, time cost may not be a reliable metric for efficiency.

\begin{table}[!ht]
	\small
	\centering
	\setlength\tabcolsep{2pt}
	\caption{Average PraPR time cost (s) \& \# patches per bug~\cite{ghanbari2019practical}.}
	\label{tab:PraPRTimeCost}
    	\begin{threeparttable}
			\begin{tabular}{l|r|r}
			\toprule
			{\bf Subjects} & {\bf \# Validated Patches} & {\bf Time cost (s)}\\
			\hline
			Chart   &  2,827.6   & 157.8   \\
			Closure &  29,849.9  & 3,027.3 \\
			\bottomrule
		\end{tabular}
		\end{threeparttable}
\end{table}

To further highlight the biases that execution time may carry, we refer to literature settings of time budgets for running APR systems: ACS~\cite{xiong2017precise} and SimFix~\cite{jiang2018shaping} are evaluated with repair time budgets of 30 minutes and 5 hours, respectively. Furthermore, in~\cite{jiang2018shaping}, assessment comparison between ACS and SimFix does not consider the bias related to the difference between the execution platforms. A comparison of performance (in terms of how many bugs each tool can fix) may, therefore, be misleading: a given bug may have been fixed by one tool because the time budget is sufficient while it cannot be fixed by the other due to lack of time.

\begin{figure}[!h]
	\centering
    \includegraphics[width=\linewidth]{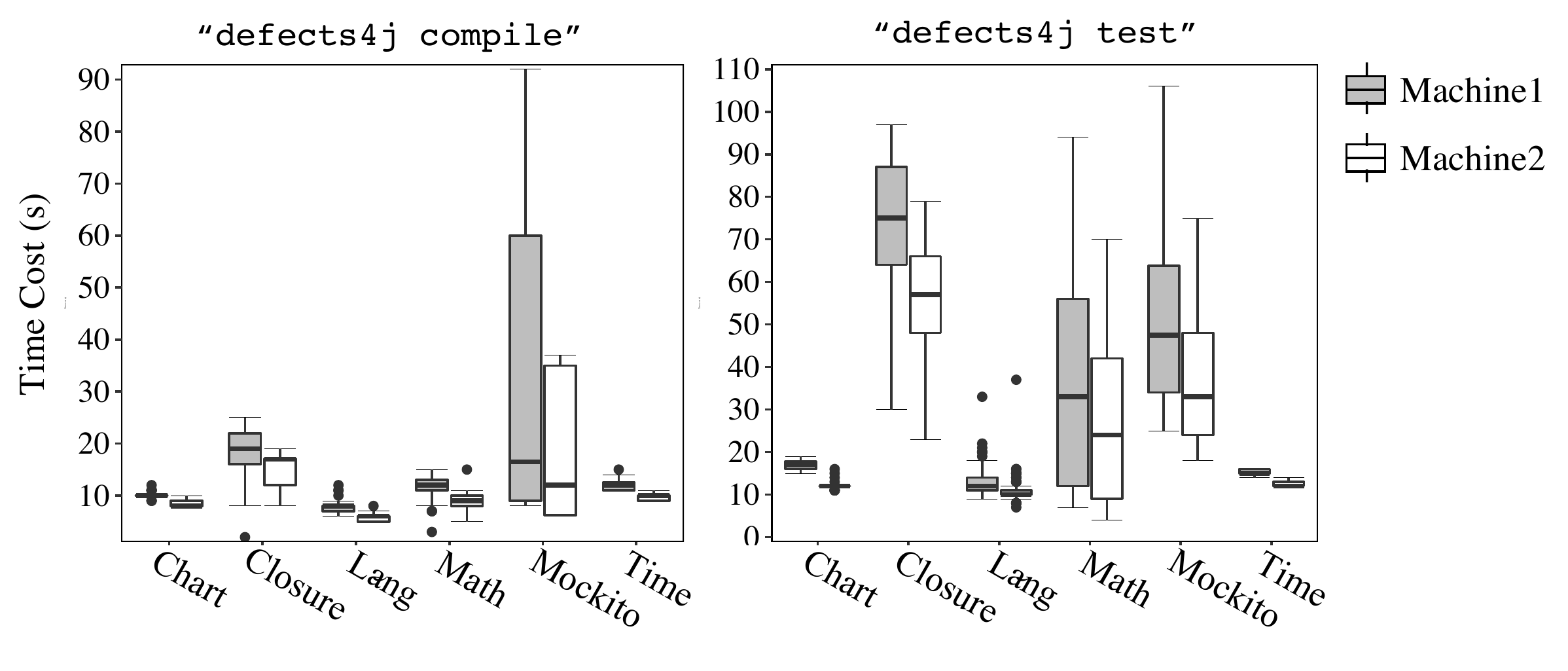}
    \vspace{-5mm}
    \caption{Distribution CPU times for compiling and testing Defects4J programs.  \\{$\bullet$ \normalfont \scriptsize {\tt Machine 1} runs OS X El Capitan 10.11.6 with 2.5 GHz Intel Core i7, 16GB 1600MHz DDR3 RAM.}\\ {$\bullet$ \normalfont \scriptsize  {\tt Machine 2} runs macOS Mojave 10.14.1 with 2.9 GHz Intel Core i9, 32 GB 2400MHz DDR4 RAM.}}
    \label{fig:compileTime}
\end{figure}

With two simple experimental runs of compiling and testing Defects4J samples,
we confirm our concerns: time budgets could introduce biases for different bugs. Indeed, as revealed in Figure~\ref{fig:compileTime}, different machine configurations may lead to drastically divergent compiling and testing time: irrespectively of projects. The Mann–Whitney–Wilcoxon tests~\cite{mann1947test,wilcoxon1945individual} confirm that the first machine consumes statistically significantly more CPU time than the second machine either for compilation or for testing Defects4J buggy programs.
These results definitively suggest that time cost is not a reliable metric to enable reproducible and comparable experiments on the efficiency of program repair.

Instead, we propose to rely on the metric of {\bf number of generated patch candidates}, which should be intrinsic to the approach and agnostic of machine configuration variabilities.

\section{Study Design}
\label{sec:exp}

This section presents the design details of this empirical study.

\vspace{-1mm}
\subsection{Research Questions}

Overall, our investigation into the efficiency of test-based APR systems seeks answers for the following research questions (RQs):
\begin{enumerate}[leftmargin=*]
  \item {\bf RQ1. Repairability across time}: We first revisit the classical
  performance criterion of APR systems, which is about the repairability (i.e.,
  effectiveness): {\em how many bugs can be fixed by test suite
  based repair approaches?} Our investigation goes beyond previous studies in
  the literature by (i) systematically assessing a large range of repair systems
  under the same configurations (see Section~\ref{sec:config}); and (ii)
  exploring not only plausibility but also the correctness of patches (see
  Section~\ref{sec:validation}). Eventually, we investigate the evolution across
  time of effectiveness to better discuss the need for revisiting efficiency as
  an important complementary performance criterion.
  \item {\bf RQ2. Patch generation efficiency}: Based on the experimental
  outputs of benchmarking repair systems in RQ1, we can investigate the
  efficiency of test-based repair: {\em how many patch candidates are generated
  and checked before fixing a given bug?} Although program repair is often regarded as a background/offline task, efficiency remains critical since resource budgets are limited. Therefore, efficiency may have adverse effects on the adoption of the repair system and even on its effectiveness. In this RQ, we
  extensively review two cases of invalid patches whose generation may undermine
  efficiency: nonsensical and in-plausible patches (see Section~\ref{sec:npc}).
	\item {\bf RQ3. Fault Localization noise impact on efficiency}: Finally, given
	that fault localization is known to provide noisy inputs to repair, we
	investigate its impact on efficiency to highlight repair directions
	for mitigations. Mainly, we question {\em whether some repair strategies are
	more or less resilient to repair attempts on wrong code locations}. Our study
	differs from recent work~\cite{liu2019you} in the literature, which explores 
	the bias of fault localization on repairability with only one repair system.
\end{enumerate}

\subsection{Subject Selection}
\label{sec:tools}
Our study focuses on APR systems targeting Java programs. Java is indeed today the most targeted language in the community of program repair. Furthermore, a well-formed dataset of real-world Java program bugs is available, with the necessary tool support to readily compile and execute programs. Although we initially planned to consider all repair approaches proposed in the last decade, we were limited by the fact that many APR tools are not open-source or even publicly available.

In the end, APR systems considered for our study are systematically selected based on the following criteria:
\begin{enumerate}[leftmargin=*]
    \item {\em Availability}: our study involves the
execution of APR tools, thus APR approaches without publicly available tools are excluded.
    \item {\em Executability}: some APR approaches provide publicly available tools,
which however cannot be executed as-is for diverse issues (e.g., ssFix~\cite{xin2017leveraging} failed to
execute because of an online connection to a private search engine fails). We exclude such approaches from the study.
    \item {\em Configurability}: to limit biases, we need to configure the different tools to use the same input information (e.g., fault localization details). We, therefore, exclude APR approaches whose tools cannot be readily configured. For example, HDRepair~\cite{le2016history} implementation is tied to an assumption that exact information on the faulty method is first available.
    \item {\em Standalone}: finally, our selection ensures that we focus on APR approaches where the tools can be run if provided with Java program source code and the available test suite. Therefore, any tool that would require extra data is excluded (e.g., LSRepair~\cite{liu2018lsrepair} requires run-time code search over Github repositories).
\end{enumerate}

We consider two sources of information to identify Java APR tools: the community-led {\em\href{http://program-repair.org/}{program-repair.org}} website and the living review of APR by Monperrus~\cite{monperrus2018living}. As of July 2019, 31 APR tools were targeting Java programs listed in the literature. After systematically examining these tools, 16 are found to satisfy our criteria and are therefore finally selected.
Table~\ref{tab:APRTools} enumerates all Java-based APR tools and provides arguments for rejection/consideration. We categorize them into three main categories: heuristic-based~\cite{le2019automated}, constraint-based~\cite{le2019automated},
and template-based~\cite{kim2013automatic} repair approaches.

\vspace{-1mm}
\paragraph{Heuristic-based repair approaches}
These approaches construct and iterate over a search space of syntactic program modifications~\cite{le2019automated}. Associated tools  include {jGenProg}~\cite{martinez2016astor}, {GenProg-A}~\cite{yuan2018arja}, ARJA~\cite{yuan2018arja}, {RSRepair-A}~\cite{yuan2018arja}, {SimFix}~\cite{jiang2018shaping}, {jKali}~\cite{martinez2016astor}, {Kali-A}~\cite{yuan2018arja}, and {jMutRepair}~\cite{martinez2016astor}. jGenProg and GenProg-A are Java implementations of
GenProg~\cite{weimer2009automatically}, which generates patches by searching
donor code from existing code with the genetic programming method. ARJA is also
a genetic programming approach to optimizing the exploration of the search space
by combining three different approaches. RSRepair-A is a Java implementation of
RSRepair~\cite{qi2014strength}, a Random-Search-based Repair tool, which tries
to repair faulty programs with the same mutation operations as GenProg but uses
random search, rather than genetic programming, to guide the patch generation
process. SimFix utilizes code change operations from existing patches and similar code to build two search
spaces, of which intersection is further used to search fix ingredients for
repairing bugs. jKali and Kali-A are Java implementations of
Kali~\cite{qi2015analysis} that fixes bugs with three actions: removal of
statements, modification of if conditions to true/false, and insertion of return
statements. jMutRepair implements the mutation-based repair
approach~\cite{debroy2010using} for Java programs, with three kinds of mutation
operators (i.e., relational, logical and unary) to fix buggy {\em if-condition}
statements.

\begin{table}[!t]
	\centering
	\scriptsize
	\caption{Included and excluded APR tools for our study.}
	\label{tab:APRTools}
	\begin{threeparttable}
		\begin{tabular}{l|l|c}
			\toprule
			{\bf Selected} & {\bf Reason} & {\bf APR Tools for Java Programs}  \\
			\hline
			No  & Not public & \makecell[c]{PAR~\cite{kim2013automatic}, xPAR~\cite{le2016history}, JFix/S3~\cite{le2017s3}, ELIXIR~\cite{saha2017elixir}, \\ Hercules~\cite{saha2019harnessing}, SOFix~\cite{liu2018mining}, CapGen~\cite{wen2018context}, PraPR$^\$$~\cite{ghanbari2019practical}.}\\
			\hline
			No & \makecell[l]{Faulty method\\required} & HDRepair~\cite{le2016history}, JAID~\cite{chen2017contract}, SketchFix~\cite{hua2018towards}.\\
			\hline
			No & Other & \makecell[c]{LSRepair$^\ast$~\cite{liu2018lsrepair}, ssFix$^\star$~\cite{xin2017leveraging},  DeepRepair$^\dagger$~\cite{white2019sorting}, NPEFix$^\ddagger$~\cite{durieux2017dynamic}.}\\
			\hline
			Yes & \makecell[l]{Open-source \\ \& working} & \makecell[c]{jGenProg~\cite{martinez2016astor}, jKali~\cite{martinez2016astor}, jMutRepair~\cite{martinez2016astor}, Cardumen~\cite{martinez2018ultra}, \\DynaMoth~\cite{durieux2016dynamoth}, Nopol~\cite{xuan2017nopol},  ACS~\cite{xiong2017precise}, SimFix~\cite{jiang2018shaping}, \\kPAR~\cite{liu2019you}, FixMiner~\cite{koyuncu2020fixminer}, AVATAR~\cite{liu2019avatar}, TBar~\cite{liu2019tbar}, \\ ARJA~\cite{yuan2018arja}, GenProg-A~\cite{yuan2018arja}, Kali-A~\cite{yuan2018arja}, RSRepair-A~\cite{yuan2018arja}.}\\
			\bottomrule
		\end{tabular}
		{$^\$$PraPR was not available before August 2019. $^\ast$LSRepair relies on the data from the run-time GitHub repositories and needs a private deep learning model~\cite{liu2018mining2} and an online code search engine~\cite{kim2018facoy} to search syntactically- or semantically-similar code, which would be biased to assess its repair efficiency.
		$^\star$ssFix fails to execute as it relies on a private code search engine that is failed to connect.
		$^\dagger$DeepRepair is not working, thus it is not selected.
		$^\ddagger$NPEfix is not selected as it does not use any fault localization technique.}
	\end{threeparttable}
\end{table}

\vspace{-2mm}
\paragraph{Constraint-based repair approaches}
These approaches generally focus on fixing a single conditional
expression that is more prone to defects than other types of program elements.
{Nopol}~\cite{xuan2017nopol}, {DynaMoth}~\cite{durieux2016dynamoth}
{ACS}~\cite{xiong2017precise}, and {Cardumen}~\cite{martinez2018ultra}
are dedicated to repairing buggy {\tt if} conditions
and to adding missing {\tt if} preconditions. Nopol relies on an SMT solver to solve the
condition synthesis problem. DynaMoth leverages the runtime context, which is a collection of
variable and method calls, to synthesize conditional expressions.
ACS is proposed to refine the ranking of ingredients
for condition synthesis.
Cardumen repairs bugs by synthesizing
patch candidates at the level of expressions with its mined templates from the
program under repair to replace the buggy expression.

\vspace{-2mm}
\paragraph{Template-based repair approaches}
These approaches are also often referred to as pattern-based and include
kPAR~\cite{liu2019you}, AVATAR~\cite{liu2019avatar}, {FixMiner}~\cite{koyuncu2020fixminer} and TBar~\cite{liu2019tbar}. kPAR is the Java
implementation of PAR~\cite{kim2013automatic} that repairs bugs with fix patterns manually summarized
from human-written patches. FixMiner automatically mines fix patterns from the
code repository for patch generation. AVATAR relies on the fix patterns for
static analysis violations. TBar combines diverse fix patterns collected from
the literature.

Note that, technically, template-based repair approaches can be viewed as
heuristics-based approaches. In this study, however, we separate them in their
category to highlight their specificity. Finally, there exist some repair
approaches that are enhanced by machine learning techniques. Le Goues et
al.~\cite{le2019automated} refer to them as  {\em learning-based} repair
approaches. One example of such approaches is the Prophet tool by Long and
Rinard~\cite{long2016automatic}: it learns from a corpus of code a model of
correct code, which indicates how likely a given piece of code is w.r.t. the code corpus. Our criteria of subject selection however excluded all
learning-based repair as they are generally not ``standalone''.

\notez{Our study considers the most diverse set of repair tools in the
literature for a systematic assessment of APR. Notably, we cover different
categories of repair approaches, while the previous record for a large scale
study, which is held by Durieux et al.~\cite{durieux2019empirical} on APR
benchmark overfitting, did not consider the most widespread template-based
tools. Furthermore, their study did not include ACS and SimFix from the current
state-of-the-art in Java APR. }

\subsection{Experiment Settings}
We now overview the inputs (i.e., buggy programs and fault localization information) and the validation process used in our study.

\vspace{-1mm}
\subsubsection{Defect Benchmark}
The APR literature includes several benchmarks~\cite{saha2018bugs,just2014defects4j,madeiral2019bears,
kim2013automatic}. In recent work, Durieux et al. showed that APR system may overfit the study benchmarks in terms of repairability. Since our objective is on efficiency, we focus on a single commonly used benchmark in the literature. We consider Defects4J~\cite{just2014defects4j} as it has been widely
employed to assess approaches~\cite{le2016history,liu2018lsrepair,jiang2018shaping,wen2018context}, or to conduct various APR studies~\cite{long2016analysis,wen2019exploring,wang2019attention,sobreira2018dissection},
as well as other software engineering
research~\cite{le2016learning,qi2013using,allamanis2018survey,pearson2017evaluating}.
Defects4J consists of \nbugs bugs across six Java open source projects. Its
dissection information~\cite{sobreira2018dissection} shows that the dataset contains
a diversity of bug types. Our experiments thus consist of running each selected APR tool to generate patches in an attempt to fix each Defects4J bug. Overall, our experiments led to 347,603 repair attempts (each attempt requiring program compilation and testing against the test suite).

\vspace{-1mm}
\subsubsection{Fault Localization}
\label{sec:config}
As reported by Liu et al.~\cite{liu2019you}, repair performance of APR tools
could be biased by fault localization settings.
To minimize such potential bias, we take on the challenge and implementation effort to re-configure all APR tools so that they are using the same fault localization information for each Defects4J bug.
In our experiments, we employ the latest release of GZoltar v1.7.2, an on-hand test automation framework. Note that early versions of this tool were widely used in the APR community~\cite{martinez2016astor,xiong2017precise,jiang2018shaping,wen2018context}. However, Liu et al. revealed that the new version yields better results in the context of program repair~\cite{liu2019you}. For sorting suspicious statements, we use the {\tt Ochiai}\cite{abreu2007accuracy} ranking metric. Eventually, APR tools are fed with a ranked list of suspicious source code statements
that should be changed within the buggy program to repair it.

\vspace{-1mm}
\subsubsection{Patch Validation}
\label{sec:validation}

Patch validation is performed by APR systems based on the execution outcome of
regression and bug-triggering test cases, i.e., test cases that are passed by
the buggy program and those that, because they are not passed, reveal the
existence of a bug. If a patch candidate can make the revised buggy program pass
the entire test suite successfully, it is considered as a valid patch. Such a
patch, however, could be incorrect if it is just overfitting the test
suite~\cite{qi2015analysis,xiong2018identifying}. Thus, the community has
adopted the terminology of {\em plausible}~\cite{qi2015analysis} patches to
refer to patches that pass all test cases.

In recent literature, following the criticism on overfitting,
researchers are shifting towards investigating {\em
correctness}~\cite{xiong2018identifying,le2019reliability}. So far, this has
been a manual effort based on a recurrent criterion: {\em a plausible patch is
considered as correct when it is semantically similar to the developer's patch
in the benchmark}. Unfortunately, the scope of semantics for APR is not
explicitly defined as it is subjective.

We propose in this work to provide a first attempt of explicitly determining semantic similarity among patches. Our objective is to reduce the threat of subjectivity and enable reproducible experiments. To that end, we call on the community and consider labels of patches within APR research artifacts. We manually revisit patches that are generated by APR tools and which researchers have considered as correct in the literature. The objective is to unveil the implicit rules that researchers use to make the decisions on correctness. We find that there are broadly two scenarios when comparing a generated patch against the developer-provided patch:
\begin{enumerate}[leftmargin=*]
    \item {\bf Identical patches}: in this case, the two patches are exactly identical, excluding variations in whitespace, layout, and comments.
    \item {\bf Semantically-similar patches}: in this case, the patches are not identical, although developers regard that they have the same effect on the program behavior. In Table~\ref{tab:incorrect} we summarize a taxonomy of correctness decision based on our study of patches labeled as correct by the research community. This taxonomy is based on the patches generated by ACS, SimFix, AVATAR, FixMiner, kPAR, and TBar whose authors investigated correctness and provided their manually labeled patches as research artifacts.
\end{enumerate}

\begin{table}[!t]
	\centering
	\scriptsize
	\caption{Example rules that the community applies to confirm  semantic similarity between tool-generated and developer-provided patches.}
	\label{tab:incorrect}
	\resizebox{1\linewidth}{!}{
	\begin{threeparttable}
		\begin{tabular}{c|L{22mm}|l}
			\toprule
			{\bf Rule ID} & {\bf Rule description}  & {\bf Illustrations} \\
			\hline
			\multirow{2}{*}{\bf R1} & {Different fields with the same value (or alias)} & 
\begin{lstlisting}[language=diff]
-    return cAvailableLocaleSet.contains(locale);
+    return availableLocaleList().contains(locale);
\end{lstlisting}\\\cline{2-2}
			 & {e.g., AVATAR$\rightarrow$Chart-7} &\cellcolor{lightgray!50}{
\begin{lstlisting}[language=diff]
+    return cAvailableLocaleList.contains(locale);
\end{lstlisting}}\\
			 \hline

			\multirow{2}{*}{\bf R2} & {Same exception but different messages} & 
\begin{lstlisting}[language=diff]
+    throw new NumberFormatException(str + 
                  (@\textcolor{codegreen}{" is not a valid number.");}@)
\end{lstlisting} \\\cline{2-2}
			& {e.g., ACS$\rightarrow$Time-15} & \cellcolor{lightgray!50}{
\begin{lstlisting}[language=diff]
+    throw new NumberFormatException();
\end{lstlisting}}\\
			\hline

			\multirow{2}{*}{\bf R3} & Variable initialization with {\tt new} rather than a default value & 
\begin{lstlisting}[language=diff]
+    if (str == null) str = "";
\end{lstlisting}\\\cline{2-2}
			&  e.g., TBar$\rightarrow$Lang-47 & \cellcolor{lightgray!50}{
\begin{lstlisting}[language=diff]
+    if (str == null) str = new String();
\end{lstlisting}}\\
			\hline

			\multirow{3}{*}{\bf R4} & {\tt if} statement instead  & 
\begin{lstlisting}[language=diff]
+    classes[i] = array[i] == null ? null : array[i].getClass();
\end{lstlisting} \\
			& of a ternary operator & \cellcolor{lightgray!50}{
\begin{lstlisting}[language=diff]
+    if (array[i] == null) continue;
\end{lstlisting}}\\\cline{2-2}
			& e.g., TBar$\rightarrow$Lang-33 & \cellcolor{lightgray!50}{
\begin{lstlisting}[language=diff]
+    classes[i] = array[i].getClass();
\end{lstlisting}} \\
			\hline

			\multirow{3}{*}{\bf R5} & Unrolling a method & 
\begin{lstlisting}[language=diff]
-    this.elitismRate = elitismRate;
+    setElitismRate(elitismRate);
\end{lstlisting} \\
			 &  & \cellcolor{lightgray!50}{
\begin{lstlisting}[language=diff]
+    if (elitismRate>(double)1.0){throw ...;}
\end{lstlisting}}\\\cline{2-2}
			 & e.g., ACS$\rightarrow$Math-35 & \cellcolor{lightgray!50}{
\begin{lstlisting}[language=diff]
+    if (elitismRate<(double)0.0){throw ...;}
\end{lstlisting}}\\
			\hline
			\multirow{2}{*}{\bf R6} & Replacing a value without a side effect
			& 
\begin{lstlisting}[language=diff]
-    int g = (int) ((value - this.lowerBound) / (this.upperBound             
+    int g = (int) ((v - this.lowerBound) / (this.upperBound 
\end{lstlisting} \\\cline{2-2}
			& \makecell[l]{e.g.,\\FixMiner$\rightarrow$Chart-24 } & \cellcolor{lightgray!50}{
\begin{lstlisting}[language=diff]
-    v = Math.min(v, this.upperBound);
+    value = Math.min(v, this.upperBound);
\end{lstlisting}}\\
			\hline

			\multirow{2}{*}{\bf R7} & Enumerating & 
\begin{lstlisting}[language=diff]
-    if (fa * fb >= 0.0 ) {            
+    if (fa * fb > 0.0 ) {
\end{lstlisting} \\\cline{2-2}
			& e.g., ACS$\rightarrow$Math\_85 & \cellcolor{lightgray!50}{
\begin{lstlisting}[language=diff]
+    if (fa * fb >= 0.0 &&!(fa * fb==0.0))
\end{lstlisting}}\\
			\hline

			{\bf R8} & Unnecessary code uncleaned  & 
\begin{lstlisting}[language=diff]
- boolean wasWhite= false;            
  for(int i= 0; i<value.length(); ++i) {            
-    if(Character.isWhitespace(c)) { ...... }            
-    wasWhite= false;
\end{lstlisting} \\\cline{2-2}
			& \makecell[l]{e.g.,\\AVATAR$\rightarrow$Lang-10} & \cellcolor{lightgray!50}{
\begin{lstlisting}[language=diff]
-    if(Character.isWhitespace(c)) { ...... }            
-    wasWhite= false;
\end{lstlisting}}\\
			\hline
			{\bf R9} & Return earlier instead of a packaged return
			& 
\begin{lstlisting}[language=diff]
-    return foundDigit && !hasExp;            
+    return foundDigit && !hasExp && !hasDecPoint;
\end{lstlisting} \\\cline{2-2}
			& e.g., ACS$\rightarrow$Lang-24 & \cellcolor{lightgray!50}{
\begin{lstlisting}[language=diff]
+    if (hasDecPoint==true){return false;} 
\end{lstlisting}}\\
			\hline

			{\bf R10} & More null checks  & 
\begin{lstlisting}[language=diff]
+    if (searchList[i] == null || replacementList[i] == null)
+    {    continue;    }
\end{lstlisting}\\\cline{2-2}
			& e.g., SimFix$\rightarrow$Lang-39 & \cellcolor{lightgray!50}{
\begin{lstlisting}[language=diff]
+    if(noMoreMatchesForReplIndex[i]||searchList[i]==null
+    ||searchList[i].length()==0||replacementList[i]==null)
+    {    continue;    }
\end{lstlisting}} \\
			\bottomrule
		\end{tabular}
		{\footnotesize We applied these rules to determine whether a plausible patch is a correct one when it is syntactically different from the patch that a developer wrote.
		In the second column, ``tool\_name$\rightarrow$bugID'' denotes that the patch generated by the tool is identified as correct.
		The patches in the grey background are generated by APR tools while the patches in the white background are patches written by the developers.}
	\end{threeparttable}
	}
\end{table}

In the remainder of this paper, for the experiments with the 16 APR tools, we
will systematically build on the rules of Table~\ref{tab:incorrect}\footnote{We enumerated
only 10 rules in this paper due to space limitation. Please visit \url{https://github.com/SerVal-DTF/APR-Efficiency} for more rules and detailed descriptions.} to label
plausible patches as correct. Thus, unless a generated patch is identical to the
developer patch, it must fall under rules R1-10 to be labeled as
correct.
Our rules are certainly
not exhaustive neither for defining semantic similarity nor for defining patch
correctness. We call on a community effort to augment these rules to
enable reproducible research.


Due to space constraints, we only detail here a single rule.
Consider rule R5: In the illustration example, the developer patch ensures that
boundaries are checked by calling a function that implements it. In contrast, a
patch generated by ACS~\cite{xiong2017precise} directly inserts the necessary
code to check the boundary. Both patches, which are not syntactically identical,
are semantically similar.

%

In the end, plausible and correct patches have the following relationship: Let $P$ and $C$ be
sets of plausible and correct patches, respectively. It always holds
$C \subseteq P$. We compute $\frac{|C|}{|P|}$ as the {\bf\em Correctness Ratio} ({\bf CR}) of generated plausible
patches that are correct.

\subsubsection{Halting Threshold}

In the APR community, it is commonly accepted that patch generation processes
are halted if a system runs out of the time budget before being able to find a
valid patch. As discussed in Section~\ref{sec:bg}, time can be a biased metric.
Therefore, in this study, we propose to halt the repair systems by setting
a threshold of repair attempts for a given bug. We set the threshold of attempts
as 10,000. This number is selected based on the reported average number
(9,696.5) of patch candidates generated by PraPR~\cite{ghanbari2019practical}
for its fixed bugs. Given that PraPR works at the mutation level and does not
require re-compilation, the number of attempts could be higher than that of other tools
and it is high enough for the 16 APR tools employed in this study.

\subsection{Terminology}

Given that {\em correct} patches are first and foremost
{\em plausible} patches, we propose in this work to use the term {\bf valid}
patches when referring to all plausible patches (including correct ones). Unless
otherwise specified, we will also refer to as {\bf  plausible} all valid patches
that have not yet been manually assessed as correct. We consciously avoid the
term {\em incorrect} since the definition of correctness in
Section~\ref{sec:validation} is sound, to some extent\footnote{The
developer-patch provided in the benchmark, which we use as ground truth, may
 be erroneous as well.}, but is not complete (i.e, there are some cases of
semantic similarity that are missed).

\subsection{Efficiency Metric: NPC}
\label{sec:npc}

As motivated in Section~\ref{sec:bg}, we employ as efficiency metric
in this study {\em the number of patch candidates} (NPC) generated by APR tools
until the first plausible patch is found. This metric was initially proposed by Qi et
al.~\cite{qi2013using} as a proxy to measure the performance of fault localization
techniques based on program repair tools.  JAID~\cite{chen2017contract} and
PraPR~\cite{ghanbari2019practical} recently used them to highlight the performance
of their approaches. Nevertheless, efficiency has not been systematically assessed before.
In this study, we further differentiate generated patches that turn out to be invalid into two groups:
\begin{enumerate}[leftmargin=*]
    \item {\bf Nonsensical patch}: Such a patch cannot even make the patched buggy program successfully compile~\cite{kim2013automatic,monperrus2014critical}.
    \item {\bf In-plausible patch}: Such a patch lets the patched buggy program successfully compile, but fails to pass some test cases in the available test suite.
\end{enumerate}
Our efficiency metric is then computed by summing the number of patches in each category:
$$NPC =NPC_{nonsensical} + NPC_{in-plausible} + NPC_{valid}$$
In practice, $NPC_{valid}==1$ since the generation of patches is halted as soon as the first valid patch is found. In this study, since we aim to investigate the repair efficiency, we focus on bugs for which the repair attempts were successfully concluded. Thus, our experimental data do not mention the cases where many patch candidates are generated but none of them was valid. We leave this investigation as a future study.

\section{Study Results}
\label{sec:eval}
We now provide experimental data as well as the key insights that are relevant to our research questions.

\vspace{-2mm}
\subsection{RQ1: Repairability Across Time}
\label{sec:rq1}

Table~\ref{tab:NFL} provides execution outcomes of 16 repair tools on the Defects4J benchmark. We count the number of bugs that are plausibly fixed by each tool implementation, and further provide the number of plausible patches that can be considered as correct following the rules of patch validation (cf. Section~\ref{sec:validation}).

\begin{table}[!h]
	\centering
	\scriptsize
	\caption{Numbers of Defects4J bugs that are correctly (plausibly) fixed by the different APR tools.}
	\label{tab:NFL}
	\resizebox{1\linewidth}{!}{
	\begin{threeparttable}
		\begin{tabular}{lcccccccc}
			\toprule
			{\bf APR Tool} & {\bf C} & {\bf Cl} & {\bf L} & {\bf M} & {\bf Mc} & {\bf T} & {\bf Total} & {\bf CR(\%)}\\
			\hline
			jGenProg   & 0 (5) & 2 (2) & 0 (2) & 3 (11)& 0 (0) & 0 (0) & 5 (20) & 25\\
			GenProg-A  & 0 (5) & 2 (15)& 0 (1) & 0 (9) & 0 (0) & 0 (0) & 2 (30) & 6.7\\
			jMutRepair & 1 (4) & 2 (5) & 0 (2) & 2 (11)& 0 (0) & 0 (0) & 5 (22) & 22.7\\
			kPAR       & 3 (13)& 2 (10)& 1 (18)& 4 (22)& 0 (0) & 0 (1) &10 ({63}) & 15.9\\
			RSRepair-A & 0 (4) & 4 (22)& 0 (3) & 0 (12)& 0 (0) & 0 (0) & 4 (41) & 10\\
			jKali      & 0 (4) & 3 (8) & 2 (4) & 1 (9) & 0 (0) & 0 (0) & 6 (25) & 24\\
			Kali-A     & 0 (6) & 2 (48)& 0 (0) & 1 (10)& 0 (1) & 0 (0) &3 ({\bf65})& 4.6\\
			DynaMoth   & 0 (6) & N/A   & 0 (2) & 1 (13)& 0 (0) & 0 (1) & 1 (22) & 4.5\\
			Nopol      & 0 (6) & N/A   & 1 (6) & 0 (18)& 0 (0) & 0 (1) & 1 (31) & 3.2\\
			ACS        & 2 (2) & 0 (0) & 3 (3) &11 (16)& 0 (0) & 1 (1) &{17} (22)& {\bf77.3}\\
			Cardumen   & 2 (4) & 0 (2) & 0 (0) & 1 (6) & 0 (0) & 0 (0) & 3 (12) & 25\\
			ARJA       & 1 (10)& 2 (29)& 0 (3) & 3 (15)& 0 (1) & 0 (0) & 6 ({58})& 10.3\\
			SimFix     & 3 (8) & 7 (19)& 5 (16)&10 (25)& 0 (0) & 0 (0) &{\bf25} ({\bf68})& {\bf36.8}\\
			FixMiner   & 5 (14)& 0 (2) &	 0 (2) & 7 (15)& 0 (0) & 0 (0) &{12} (33) & {\bf36.4}\\
			AVATAR     & 5 (12)& 7 (15)& 4 (13)& 3 (17)& 0 (0) & 0 (0) &{\bf19} (57)& 33.3\\
			TBar       & 7 (16)& 3 (12)& 6 (21)& 8 (23)& 0 (0) & 0 (0) &{\bf24} ({\bf72})& 30.8\\
			\bottomrule
		\end{tabular}
		{$^\ast$The numbers outside the parentheses indicate the bugs fixed with correct
		patches while the numbers inside parentheses indicate the number of plausible
		patches. The missing numbers are marked with N/A as we failed to change the
		fault localization input for {\tt Closure} program bugs for DynaMoth and Nopol,
		of which fault localization is tightly tied with GZoltar-0.0.1.
		``C, Cl, L, M, Mc, and T'' represent Chart, Closure, Lang, Math, Mockito and Time, respectively. The same as
		Table~\ref{tab:PFL_REPAIR}.}
	\end{threeparttable}
	}\vspace{-1mm}
\end{table}

\begin{table*}[htbp]
  \scriptsize
  \centering
  \caption{Number of overlapped fixed bugs per repair tool.}
  \label{tab:NFL_overlap}
    \resizebox{\textwidth}{!}
    {
    \begin{threeparttable}
    	\begin{tabular}{l|cccccccccccccccc}
    \toprule
          & jGenProg & GenProg-A & jMutRepair & kPAR  & RSRepair-A & jKali & Kali-A & DynaMoth & Nopol & ACS   & Cardumen & ARJA  & SimFix & FixMiner & AVATAR & TBar \\
    \midrule
    jGenProg & 5.0\% (1) & \cellcolor[rgb]{ .682,  .667,  .667}40.0\% (8) & \cellcolor[rgb]{ .459,  .443,  .443}45.0\% (9) & \cellcolor[rgb]{ .459,  .443,  .443}55.0\% (11) & \cellcolor[rgb]{ .459,  .443,  .443}45.0\% (9) & \cellcolor[rgb]{ .682,  .667,  .667}40.0\% (8) & \cellcolor[rgb]{ .682,  .667,  .667}40.0\% (8) & \cellcolor[rgb]{ .682,  .667,  .667}35.0\% (7) & \cellcolor[rgb]{ .682,  .667,  .667}25.0\% (5) & \cellcolor[rgb]{ .816,  .808,  .808}20.0\% (4) & \cellcolor[rgb]{ .816,  .808,  .808}30.0\% (6) & \cellcolor[rgb]{ .459,  .443,  .443}60.0\% (12) & \cellcolor[rgb]{ .227,  .22,  .22}\textcolor[rgb]{ 1,  1,  1}{80.0\% (16)} & \cellcolor[rgb]{ .459,  .443,  .443}45.0\% (9) & \cellcolor[rgb]{ .459,  .443,  .443}60.0\% (12) & \cellcolor[rgb]{ .051,  .051,  .051}\textcolor[rgb]{ 1,  1,  1}{85.0\% (17)} \\
    GenProg-A & \cellcolor[rgb]{ .682,  .667,  .667}26.7\% (8) & 0.0\% (0) & \cellcolor[rgb]{ .682,  .667,  .667}36.7\% (11) & \cellcolor[rgb]{ .459,  .443,  .443}46.7\% (14) & \cellcolor[rgb]{ .051,  .051,  .051}\textcolor[rgb]{ 1,  1,  1}{90.0\% (27)} & \cellcolor[rgb]{ .682,  .667,  .667}33.3\% (10) & \cellcolor[rgb]{ .227,  .22,  .22}\textcolor[rgb]{ 1,  1,  1}{80.0\% (24)} & \cellcolor[rgb]{ .682,  .667,  .667}23.3\% (7) & \cellcolor[rgb]{ .816,  .808,  .808}20.0\% (6) & \cellcolor[rgb]{ .816,  .808,  .808}16.7\% (5) & \cellcolor[rgb]{ .816,  .808,  .808}10.0\% (3) & \cellcolor[rgb]{ .051,  .051,  .051}\textcolor[rgb]{ 1,  1,  1}{96.7\% (29)} & \cellcolor[rgb]{ .682,  .667,  .667}40.0\% (12) & \cellcolor[rgb]{ .682,  .667,  .667}30.0\% (9) & \cellcolor[rgb]{ .459,  .443,  .443}43.3\% (13) & \cellcolor[rgb]{ .459,  .443,  .443}53.3\% (16) \\
    jMutRepair & \cellcolor[rgb]{ .459,  .443,  .443}40.9\% (9) & \cellcolor[rgb]{ .459,  .443,  .443}50.0\% (11) & 4.5\% (1) & \cellcolor[rgb]{ .227,  .22,  .22}\textcolor[rgb]{ 1,  1,  1}{68.2\% (15)} & \cellcolor[rgb]{ .459,  .443,  .443}50.0\% (11) & \cellcolor[rgb]{ .459,  .443,  .443}59.1\% (13) & \cellcolor[rgb]{ .459,  .443,  .443}54.4\% (12) & \cellcolor[rgb]{ .682,  .667,  .667}31.8\% (7) & \cellcolor[rgb]{ .682,  .667,  .667}22.7\% (5) & \cellcolor[rgb]{ .816,  .808,  .808}18.2\% (4) & \cellcolor[rgb]{ .816,  .808,  .808}13.6\% (3) & \cellcolor[rgb]{ .227,  .22,  .22}\textcolor[rgb]{ 1,  1,  1}{63.6\% (14)} & \cellcolor[rgb]{ .227,  .22,  .22}\textcolor[rgb]{ 1,  1,  1}{77.3\% (17)} & \cellcolor[rgb]{ .459,  .443,  .443}45.5\% (10) & \cellcolor[rgb]{ .051,  .051,  .051}\textcolor[rgb]{ 1,  1,  1}{86.4\% (19)} & \cellcolor[rgb]{ .051,  .051,  .051}\textcolor[rgb]{ 1,  1,  1}{90.9\% (20)} \\
    kPAR  & \cellcolor[rgb]{ .816,  .808,  .808}17.5\% (11) & \cellcolor[rgb]{ .682,  .667,  .667}22.2\% (14) & \cellcolor[rgb]{ .682,  .667,  .667}23.8\% (15) & 6.3\% (4) & \cellcolor[rgb]{ .682,  .667,  .667}25.4\% (16) & \cellcolor[rgb]{ .682,  .667,  .667}25.4\% (16) & \cellcolor[rgb]{ .682,  .667,  .667}25.4\% (16) & \cellcolor[rgb]{ .682,  .667,  .667}22.2\% (14) & \cellcolor[rgb]{ .682,  .667,  .667}25.4\% (16) & \cellcolor[rgb]{ .816,  .808,  .808}11.1\% (7) & \cellcolor[rgb]{ .816,  .808,  .808}7.9\% (5) & \cellcolor[rgb]{ .682,  .667,  .667}39.7\% (25) & \cellcolor[rgb]{ .459,  .443,  .443}49.2\% (31) & \cellcolor[rgb]{ .682,  .667,  .667}34.9\% (22) & \cellcolor[rgb]{ .459,  .443,  .443}57.1\% (36) & \cellcolor[rgb]{ .227,  .22,  .22}\textcolor[rgb]{ 1,  1,  1}{74.6\% (47)} \\
    RSRepair-A & \cellcolor[rgb]{ .682,  .667,  .667}22.0\% (9) & \cellcolor[rgb]{ .227,  .22,  .22}\textcolor[rgb]{ 1,  1,  1}{65.9\% (27)} & \cellcolor[rgb]{ .682,  .667,  .667}26.8\% (11) & \cellcolor[rgb]{ .682,  .667,  .667}39.0\% (16) & 2.4\% (1) & \cellcolor[rgb]{ .682,  .667,  .667}26.8\% (11) & \cellcolor[rgb]{ .227,  .22,  .22}\textcolor[rgb]{ 1,  1,  1}{75.6\% (31)} & \cellcolor[rgb]{ .816,  .808,  .808}19.5\% (8) & \cellcolor[rgb]{ .682,  .667,  .667}22.0\% (9) & \cellcolor[rgb]{ .816,  .808,  .808}12.2\% (5) & \cellcolor[rgb]{ .816,  .808,  .808}7.3\% (3) & \cellcolor[rgb]{ .051,  .051,  .051}\textcolor[rgb]{ 1,  1,  1}{85.4\% (35)} & \cellcolor[rgb]{ .682,  .667,  .667}29.3\% (12) & \cellcolor[rgb]{ .816,  .808,  .808}19.5\% (8) & \cellcolor[rgb]{ .682,  .667,  .667}39.0\% (16) & \cellcolor[rgb]{ .459,  .443,  .443}41.5\% (17) \\
    jKali & \cellcolor[rgb]{ .682,  .667,  .667}32.0\% (8) & \cellcolor[rgb]{ .682,  .667,  .667}40.0\% (10) & \cellcolor[rgb]{ .459,  .443,  .443}52.0\% (13) & \cellcolor[rgb]{ .227,  .22,  .22}\textcolor[rgb]{ 1,  1,  1}{64.0\% (16)} & \cellcolor[rgb]{ .459,  .443,  .443}44.0\% (11) & 8.0\% (2) & \cellcolor[rgb]{ .459,  .443,  .443}56.0\% (14) & \cellcolor[rgb]{ .682,  .667,  .667}40.0\% (10) & \cellcolor[rgb]{ .682,  .667,  .667}24.0\% (6) & \cellcolor[rgb]{ .816,  .808,  .808}8.0\% (2) & \cellcolor[rgb]{ .816,  .808,  .808}12.0\% (3) & \cellcolor[rgb]{ .459,  .443,  .443}56.0\% (14) & \cellcolor[rgb]{ .459,  .443,  .443}56.0\% (14) & \cellcolor[rgb]{ .816,  .808,  .808}20.0\% (5) & \cellcolor[rgb]{ .227,  .22,  .22}\textcolor[rgb]{ 1,  1,  1}{76.0\% (19)} & \cellcolor[rgb]{ .227,  .22,  .22}\textcolor[rgb]{ 1,  1,  1}{68.0\% (17)} \\
    Kali-A & \cellcolor[rgb]{ .816,  .808,  .808}12.3\% (8) & \cellcolor[rgb]{ .682,  .667,  .667}36.9\% (24) & \cellcolor[rgb]{ .816,  .808,  .808}18.5\% (12) & \cellcolor[rgb]{ .682,  .667,  .667}24.6\% (16) & \cellcolor[rgb]{ .459,  .443,  .443}47.7\% (31) & \cellcolor[rgb]{ .682,  .667,  .667}21.5\% (14) & 23.1\% (15) & \cellcolor[rgb]{ .816,  .808,  .808}13.8\% (9) & \cellcolor[rgb]{ .816,  .808,  .808}9.2\% (6) & \cellcolor[rgb]{ .816,  .808,  .808}3.1\% (2) & \cellcolor[rgb]{ .816,  .808,  .808}1.5\% (1) & \cellcolor[rgb]{ .227,  .22,  .22}\textcolor[rgb]{ 1,  1,  1}{63.1\% (41)} & \cellcolor[rgb]{ .682,  .667,  .667}21.5\% (14) & \cellcolor[rgb]{ .816,  .808,  .808}15.4\% (10) & \cellcolor[rgb]{ .682,  .667,  .667}29.2\% (19) & \cellcolor[rgb]{ .682,  .667,  .667}27.7\% (18) \\
    DynaMoth & \cellcolor[rgb]{ .682,  .667,  .667}31.8\% (7) & \cellcolor[rgb]{ .682,  .667,  .667}31.8\% (7) & \cellcolor[rgb]{ .682,  .667,  .667}31.8\% (7) & \cellcolor[rgb]{ .227,  .22,  .22}\textcolor[rgb]{ 1,  1,  1}{63.6\% (14)} & \cellcolor[rgb]{ .682,  .667,  .667}36.4\% (8) & \cellcolor[rgb]{ .459,  .443,  .443}45.5\% (10) & \cellcolor[rgb]{ .459,  .443,  .443}40.9\% (9) & 0.0\% (0) & \cellcolor[rgb]{ .459,  .443,  .443}54.5\% (12) & \cellcolor[rgb]{ .816,  .808,  .808}13.6\% (3) & \cellcolor[rgb]{ .816,  .808,  .808}9.1\% (2) & \cellcolor[rgb]{ .459,  .443,  .443}50.0\% (11) & \cellcolor[rgb]{ .459,  .443,  .443}54.5\% (12) & \cellcolor[rgb]{ .459,  .443,  .443}50.0\% (11) & \cellcolor[rgb]{ .459,  .443,  .443}54.5\% (12) & \cellcolor[rgb]{ .459,  .443,  .443}59.1\% (13) \\
    Nopol & \cellcolor[rgb]{ .816,  .808,  .808}16.1\% (5) & \cellcolor[rgb]{ .816,  .808,  .808}19.4\% (6) & \cellcolor[rgb]{ .816,  .808,  .808}16.1\% (5) & \cellcolor[rgb]{ .459,  .443,  .443}51.6\% (16) & \cellcolor[rgb]{ .682,  .667,  .667}29.0\% (9) & \cellcolor[rgb]{ .816,  .808,  .808}19.4\% (6) & \cellcolor[rgb]{ .816,  .808,  .808}19.4\% (6) & \cellcolor[rgb]{ .682,  .667,  .667}38.7\% (12) & 19.4\% (6) & \cellcolor[rgb]{ .816,  .808,  .808}12.9\% (4) & \cellcolor[rgb]{ .816,  .808,  .808}6.5\% (2) & \cellcolor[rgb]{ .682,  .667,  .667}25.8\% (8) & \cellcolor[rgb]{ .682,  .667,  .667}25.8\% (8) & \cellcolor[rgb]{ .816,  .808,  .808}19.4\% (6) & \cellcolor[rgb]{ .682,  .667,  .667}38.7\% (12) & \cellcolor[rgb]{ .682,  .667,  .667}35.5\% (11) \\
    ACS   & \cellcolor[rgb]{ .816,  .808,  .808}18.2\% (4) & \cellcolor[rgb]{ .682,  .667,  .667}22.7\% (5) & \cellcolor[rgb]{ .816,  .808,  .808}18.2\% (4) & \cellcolor[rgb]{ .682,  .667,  .667}31.8\% (7) & \cellcolor[rgb]{ .682,  .667,  .667}22.7\% (5) & \cellcolor[rgb]{ .816,  .808,  .808}9.1\% (2) & \cellcolor[rgb]{ .816,  .808,  .808}9.1\% (2) & \cellcolor[rgb]{ .816,  .808,  .808}13.6\% (3) & \cellcolor[rgb]{ .816,  .808,  .808}18.2\% (4) & 40.9\% (9) & \cellcolor[rgb]{ .816,  .808,  .808}13.6\% (3) & \cellcolor[rgb]{ .682,  .667,  .667}36.4\% (8) & \cellcolor[rgb]{ .682,  .667,  .667}22.7\% (5) & \cellcolor[rgb]{ .816,  .808,  .808}18.2\% (4) & \cellcolor[rgb]{ .682,  .667,  .667}31.8\% (7) & \cellcolor[rgb]{ .459,  .443,  .443}40.9\% (9) \\
    Cardumen & \cellcolor[rgb]{ .459,  .443,  .443}50.0\% (6) & \cellcolor[rgb]{ .682,  .667,  .667}25.0\% (3) & \cellcolor[rgb]{ .682,  .667,  .667}25.0\% (3) & \cellcolor[rgb]{ .459,  .443,  .443}41.7\% (5) & \cellcolor[rgb]{ .682,  .667,  .667}25.0\% (3) & \cellcolor[rgb]{ .682,  .667,  .667}25.0\% (3) & \cellcolor[rgb]{ .816,  .808,  .808}8.3\% (1) & \cellcolor[rgb]{ .816,  .808,  .808}16.7\% (2) & \cellcolor[rgb]{ .816,  .808,  .808}16.7\% (2) & \cellcolor[rgb]{ .682,  .667,  .667}25.0\% (3) & 8.3\% (1) & \cellcolor[rgb]{ .682,  .667,  .667}25.0\% (3) & \cellcolor[rgb]{ .459,  .443,  .443}58.3\% (7) & \cellcolor[rgb]{ .459,  .443,  .443}50.0\% (6) & \cellcolor[rgb]{ .459,  .443,  .443}50.0\% (6) & \cellcolor[rgb]{ .051,  .051,  .051}\textcolor[rgb]{ 1,  1,  1}{83.3\% (10)} \\
    ARJA  & \cellcolor[rgb]{ .682,  .667,  .667}20.7\% (12) & \cellcolor[rgb]{ .459,  .443,  .443}50.0\% (29) & \cellcolor[rgb]{ .682,  .667,  .667}24.1\% (14) & \cellcolor[rgb]{ .459,  .443,  .443}43.1\% (25) & \cellcolor[rgb]{ .227,  .22,  .22}\textcolor[rgb]{ 1,  1,  1}{60.3\% (35)} & \cellcolor[rgb]{ .682,  .667,  .667}24.1\% (14) & \cellcolor[rgb]{ .227,  .22,  .22}\textcolor[rgb]{ 1,  1,  1}{70.7\% (41)} & \cellcolor[rgb]{ .816,  .808,  .808}19.0\% (11) & \cellcolor[rgb]{ .816,  .808,  .808}13.8\% (8) & \cellcolor[rgb]{ .816,  .808,  .808}13.8\% (8) & \cellcolor[rgb]{ .816,  .808,  .808}5.2\% (3) & 6.9\% (4) & \cellcolor[rgb]{ .682,  .667,  .667}31.0\% (18) & \cellcolor[rgb]{ .682,  .667,  .667}25.9\% (15) & \cellcolor[rgb]{ .682,  .667,  .667}39.7\% (23) & \cellcolor[rgb]{ .459,  .443,  .443}43.1\% (25) \\
    SimFix & \cellcolor[rgb]{ .682,  .667,  .667}23.5\% (16) & \cellcolor[rgb]{ .816,  .808,  .808}17.6\% (12) & \cellcolor[rgb]{ .682,  .667,  .667}25.0\% (17) & \cellcolor[rgb]{ .459,  .443,  .443}45.6\% (31) & \cellcolor[rgb]{ .816,  .808,  .808}17.6\% (12) & \cellcolor[rgb]{ .682,  .667,  .667}20.6\% (14) & \cellcolor[rgb]{ .682,  .667,  .667}20.6\% (14) & \cellcolor[rgb]{ .816,  .808,  .808}17.6\% (12) & \cellcolor[rgb]{ .816,  .808,  .808}11.8\% (8) & \cellcolor[rgb]{ .816,  .808,  .808}7.4\% (5) & \cellcolor[rgb]{ .816,  .808,  .808}10.3\% (7) & \cellcolor[rgb]{ .682,  .667,  .667}26.5\% (18) & 19.1\% (13) & \cellcolor[rgb]{ .682,  .667,  .667}25.0\% (17) & \cellcolor[rgb]{ .682,  .667,  .667}39.7\% (27) & \cellcolor[rgb]{ .459,  .443,  .443}58.8\% (40) \\
    FixMiner & \cellcolor[rgb]{ .682,  .667,  .667}27.3\% (9) & \cellcolor[rgb]{ .682,  .667,  .667}27.3\% (9) & \cellcolor[rgb]{ .682,  .667,  .667}30.3\% (10) & \cellcolor[rgb]{ .227,  .22,  .22}\textcolor[rgb]{ 1,  1,  1}{66.7\% (22)} & \cellcolor[rgb]{ .682,  .667,  .667}24.2\% (8) & \cellcolor[rgb]{ .816,  .808,  .808}15.2\% (5) & \cellcolor[rgb]{ .682,  .667,  .667}30.3\% (10) & \cellcolor[rgb]{ .682,  .667,  .667}33.3\% (11) & \cellcolor[rgb]{ .816,  .808,  .808}18.2\% (6) & \cellcolor[rgb]{ .816,  .808,  .808}12.1\% (4) & \cellcolor[rgb]{ .816,  .808,  .808}18.2\% (6) & \cellcolor[rgb]{ .459,  .443,  .443}45.5\% (15) & \cellcolor[rgb]{ .459,  .443,  .443}51.5\% (17) & 9.1\% (3) & \cellcolor[rgb]{ .459,  .443,  .443}54.5\% (18) & \cellcolor[rgb]{ .227,  .22,  .22}\textcolor[rgb]{ 1,  1,  1}{75.8\% (25)} \\
    AVATAR & \cellcolor[rgb]{ .682,  .667,  .667}21.1\% (12) & \cellcolor[rgb]{ .682,  .667,  .667}22.8\% (13) & \cellcolor[rgb]{ .682,  .667,  .667}33.3\% (19) & \cellcolor[rgb]{ .227,  .22,  .22}\textcolor[rgb]{ 1,  1,  1}{63.2\% (36)} & \cellcolor[rgb]{ .682,  .667,  .667}28.1\% (16) & \cellcolor[rgb]{ .682,  .667,  .667}33.3\% (19) & \cellcolor[rgb]{ .682,  .667,  .667}33.3\% (19) & \cellcolor[rgb]{ .682,  .667,  .667}21.1\% (12) & \cellcolor[rgb]{ .682,  .667,  .667}21.1\% (12) & \cellcolor[rgb]{ .816,  .808,  .808}12.3\% (7) & \cellcolor[rgb]{ .816,  .808,  .808}10.5\% (6) & \cellcolor[rgb]{ .459,  .443,  .443}40.4\% (23) & \cellcolor[rgb]{ .459,  .443,  .443}47.4\% (27) & \cellcolor[rgb]{ .682,  .667,  .667}31.6\% (18) & 5.3\% (3) & \cellcolor[rgb]{ .227,  .22,  .22}\textcolor[rgb]{ 1,  1,  1}{78.9\% (45)} \\
    TBar  & \cellcolor[rgb]{ .682,  .667,  .667}23.6\% (17) & \cellcolor[rgb]{ .682,  .667,  .667}22.2\% (16) & \cellcolor[rgb]{ .682,  .667,  .667}27.8\% (20) & \cellcolor[rgb]{ .227,  .22,  .22}\textcolor[rgb]{ 1,  1,  1}{65.3\% (47)} & \cellcolor[rgb]{ .682,  .667,  .667}23.6\% (17) & \cellcolor[rgb]{ .682,  .667,  .667}23.6\% (17) & \cellcolor[rgb]{ .682,  .667,  .667}25.0\% (18) & \cellcolor[rgb]{ .816,  .808,  .808}18.1\% (13) & \cellcolor[rgb]{ .816,  .808,  .808}15.3\% (11) & \cellcolor[rgb]{ .816,  .808,  .808}12.5\% (9) & \cellcolor[rgb]{ .816,  .808,  .808}13.9\% (10) & \cellcolor[rgb]{ .682,  .667,  .667}34.7\% (25) & \cellcolor[rgb]{ .459,  .443,  .443}55.6\% (40) & \cellcolor[rgb]{ .682,  .667,  .667}34.7\% (25) & \cellcolor[rgb]{ .227,  .22,  .22}\textcolor[rgb]{ 1,  1,  1}{62.5\% (45)} & 5.6\% (4) \\
    \bottomrule
    \end{tabular}
    {\footnotesize The intersection of tool X (row) and tool Y (column) contains the percentage of bugs fixed by X which are also fixed by Y.
    For instance, 40\% of the bugs fixed by jGenProg (row 1) are also fixed by GenProg-A (column 2). On the contrary, 26.7\% of the bugs fixed by GenProg-A (row 2) are also fixed by jGenProg (column 1). While the diagonal cells present the number of bugs exclusively fixed by each repair tool.}
  \label{tab:addlabel}%
    \end{threeparttable}
    }%
    \vspace{1.5mm}
\end{table*}%


$\bullet$ [{\em Template-based repair tools are the most effective.}]
We observe that kPAR, FixMiner, AVATAR and TBar, which are template-based repair tools, present better repair performance than other tools in terms of the number of fixed bugs. The state-of-the-art, SimFix, also performs among the top. Note that, although it is classified as heuristics-based, and does not use templates explicitly, it performs transformations based on similar changes, and thus has been presented in previous studies~\cite{liu2019tbar} as template-based.

$\bullet$ [{\em Patch ordering strategies are necessary to increase the likelihood of hitting correct patches.}] Among the 16 repair tools, ACS exhibits the highest ratio of plausible patches that are found to be correct. This experimental finding confirms the strategy used by the authors to increase ``precision''\footnote{Precision is the terminology employed by its authors to refer to the ratio of correct patches to plausible patches.} in patch generation: these are dependency-based ordering, document analysis, and predicate mining.

\begin{figure}[!t]
	\centering
	\includegraphics[width=\linewidth]{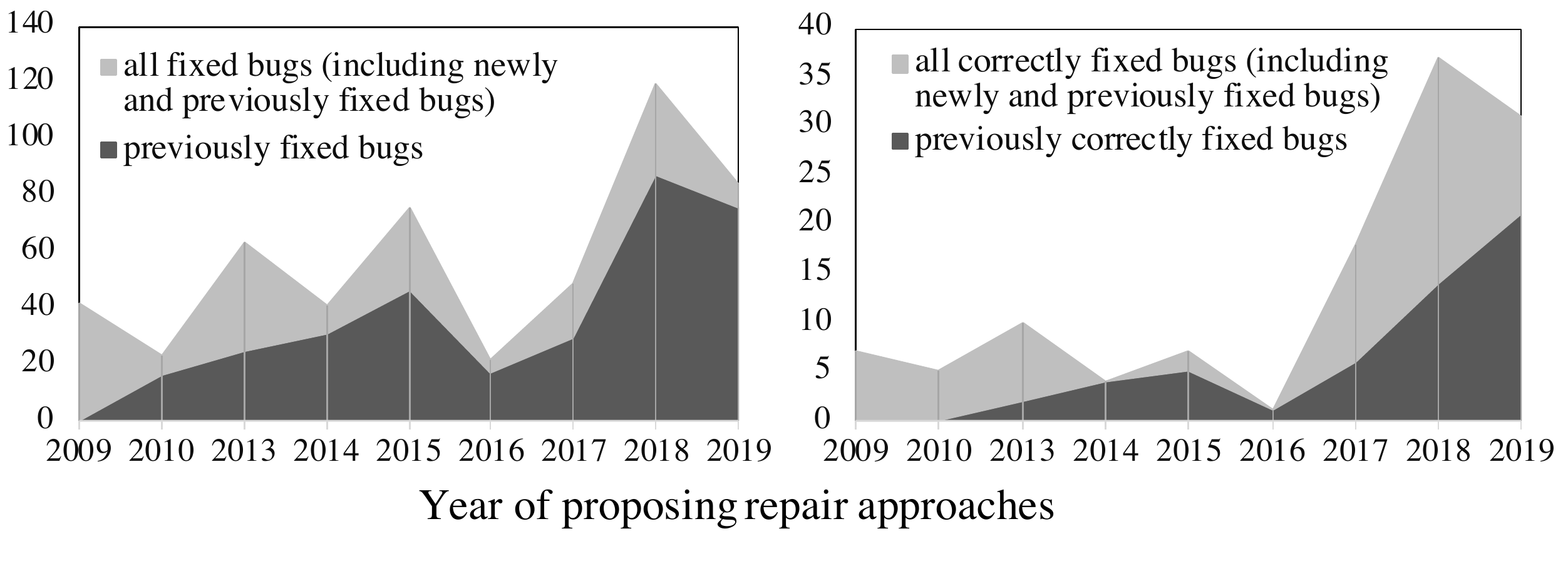}
	\vspace{-5mm}
	\caption{Evolution of the number of fixed bugs across time.}
  \label{fig:Year-fixed-bugs}
\end{figure}

$\bullet$ [{\em Through time, repair tools tend to subsume their predecessors in terms of which bugs are fixed.}]
Table~\ref{tab:NFL_overlap} provides statistics on the percentage of fixed bugs that are overlapping between two repair tools. In this table, the tools in column headers and row headers are ordered chronologically concerning the date of approach publication. Note that jGenProg ranked based on the GenProg publication year although the tool itself was implemented years later.
We note that the upper-right side of the table is relatively darker than the rest: the percentages of overlapping are higher for these cells. These results suggest that, overall, the bugs that are fixed by earlier tools are also generally covered by more recent tools. Besides, evolution trends presented in Figure~\ref{fig:Year-fixed-bugs} show that, although the number of bugs that are fixed by the different tools over the years is increasing, the number of new bugs is increasing with small increments. This result suggests that the strategies implemented in new approaches tend to have similar outcomes as merging past techniques to cover previous bug sets that were fixed each via different approaches.

\begin{figure}[!t]
    \centering
    \includegraphics[width=\linewidth]{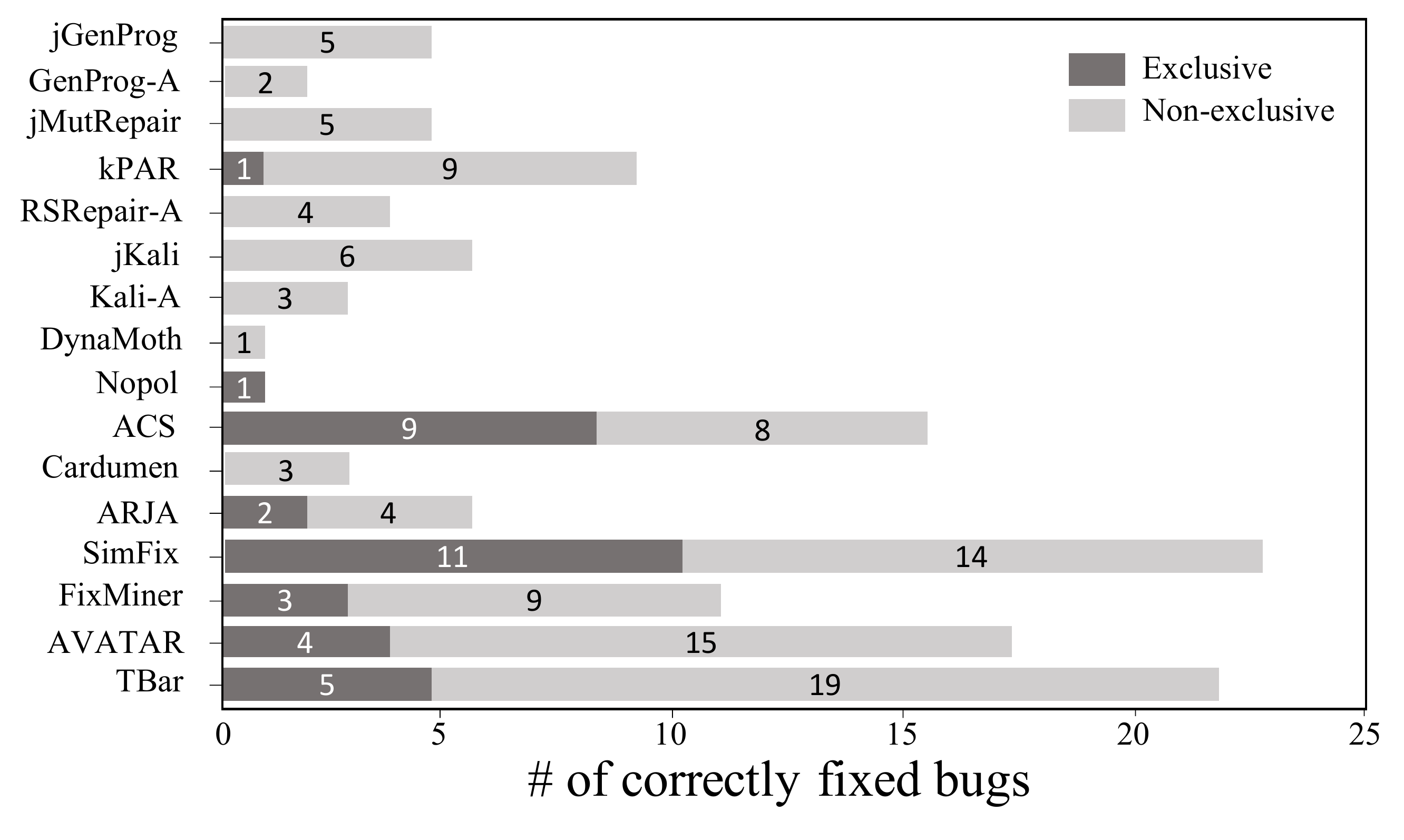}
    \vspace{-5.5mm}
    \caption{Repairing exclusivity of each APR tool (correct patches).}
    \label{fig:NFL-correct-overlap}
\end{figure}

$\bullet$ [{\em Recent APR tools tend to correctly fix more bugs than their predecessors.}]
In the right part of Figure~\ref{fig:Year-fixed-bugs}, a visible breakthrough is the sharp increase of the light grey area indicating that recent tools increasingly correctly fix bugs which have not been fixed by previous tools. 
We further summarize in Figure~\ref{fig:NFL-correct-overlap} the number of bugs that each tool can correctly fix exclusively or not.
SimFix, ACS, AVATAR, and TBar are leading repair tools that generate correct fixes for more bugs. In contrast, jGenProg, GenProg-A, jMutRepair, RSRepair-A, jKali, Kali-A, DynaMoth, and Cardumen do not correctly fix any Defects4J bug that is not also correctly fixed by another tool.

$\bullet$ [{\em Implementation details can make a difference.}]
Finally, we observe that Java-targeted implementations of GenProg (i.e, jGenProg and GenProg-A) and Kali (i.e., jKali and Kali-A) by different research groups yield diverging repair performance on the same benchmark.

\notez{Overall the systematic study of repairability of APR tools across time reveals that (1) recent tools tend to fix more bugs than their predecessors; (2)  each newly-proposed repair tool however plausibly fix few bugs that were not fixed by other tools; (3) more bugs can be correctly-fixed by lately-proposed APR tools; and (4) template-based repair tools are the most effective to eventually produce plausible patches. It thus remains unclear whether the strategies proposed by record-setting tools are improving the state-of-the-art of patch generation. We propose to focus on efficiency as a complementary metric to assess performance gains.}

\subsection{RQ2: Patch Generation Efficiency}
\label{sec:rq2}

Following our motivation argument in Section~\ref{sec:bg},
we use the $NPC$ scores (i.e., number of generated patch candidates that are checked \underline{until a valid patch is found}) to measure repair efficiency of APR tools. For each tool, the results focus on Defects4J bugs that are fixed (i.e., a valid patch was eventually found).  Indeed, through efficiency, we attempt to {\bf measure the ability of the APR tool to avoid wasting computing resource, time and energy in patch validation towards generating a valid patch}.

\begin{figure}[!ht]
    \centering
    \includegraphics[width=\linewidth]{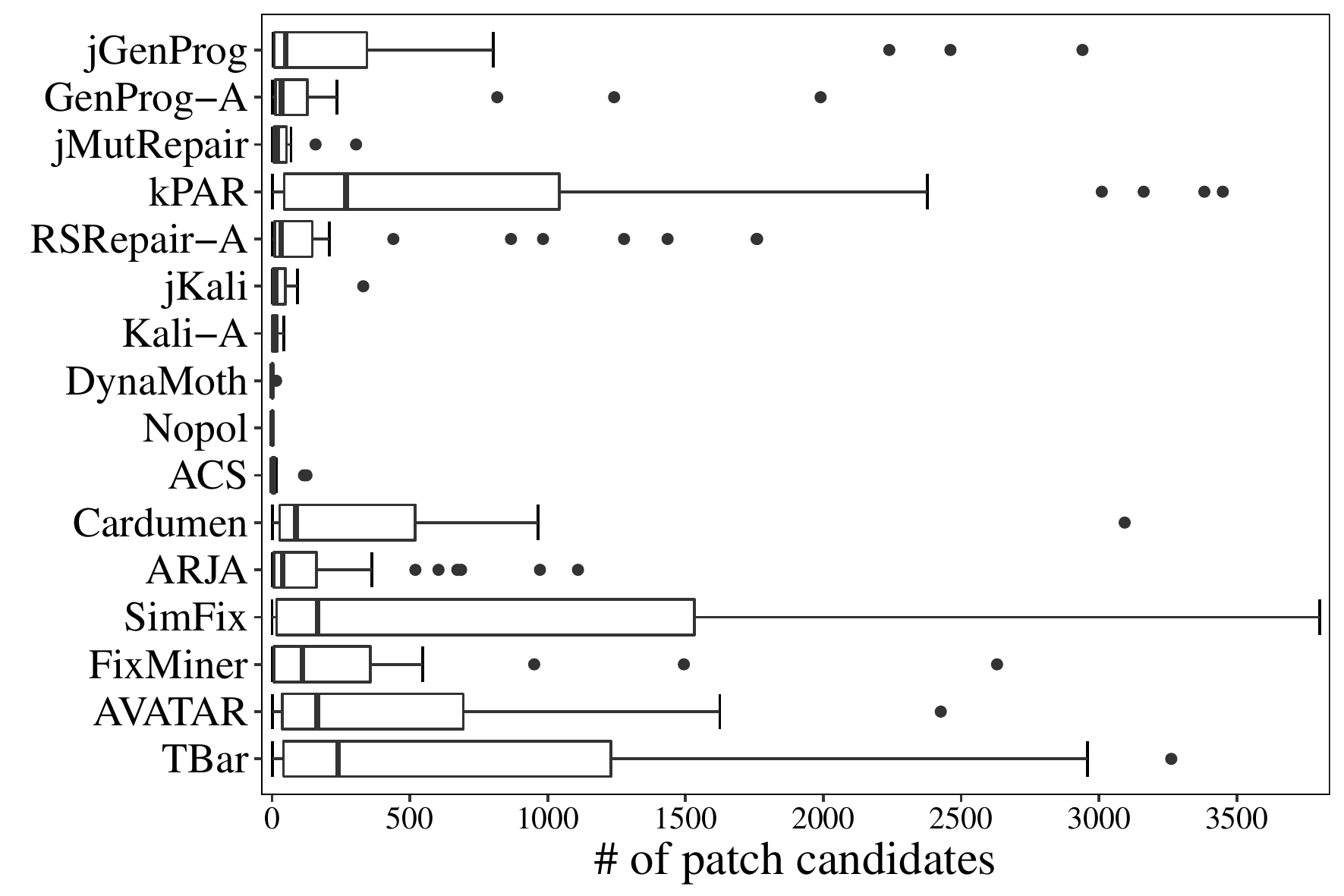}
    \caption{The distribution of NPC scores for 16 APR tools.}
    \label{fig:NFL-NPC1}
\end{figure}

Figure~\ref{fig:NFL-NPC1} overviews the general distributions of $NPC$ scores of the 16 repair tools on the Defects4J benchmark.
For all tools, the median $NPC$ is lower than 250 patch candidates.
However, the distribution spread among bugs is not only significant for several (8 out of 16) tools but also varies across tools.

$\bullet$[{\em Efficiency is not yet a widely-valued performance target.}] SimFix, TBar and kPAR exhibit the highest $NPC$ scores which can go beyond 1,000 patch candidates for some bugs.
Correlating this data with repairability findings (Section~\ref{sec:rq1}), we note that tools with highest repairability scores also have the highest $NPC$ scores (hence, {\em lower efficiency}). In particular, we note that APR approaches, which rely on change patterns (i.e., standard template-based tools) or heuristically search for donor code based on code similarity (e.g., SimFix), produce the largest number of patch candidates. They are \underline{effective} since they end-up finding a valid patch, but they are \underline{not efficient} as they generate too many patches (comparing against other approaches) for repair attempts. On the other hand, constraint-based APR tools (e.g., ACS) have the lowest $NPC$ scores.
There is, therefore, an insight that constraint-solving and synthesis strategies, although they might require more computing effort to traverse the search space, eventually yield patches which waste less resource during test-based validation.

\begin{figure}[!t]
    \centering
    \includegraphics[width=\linewidth]{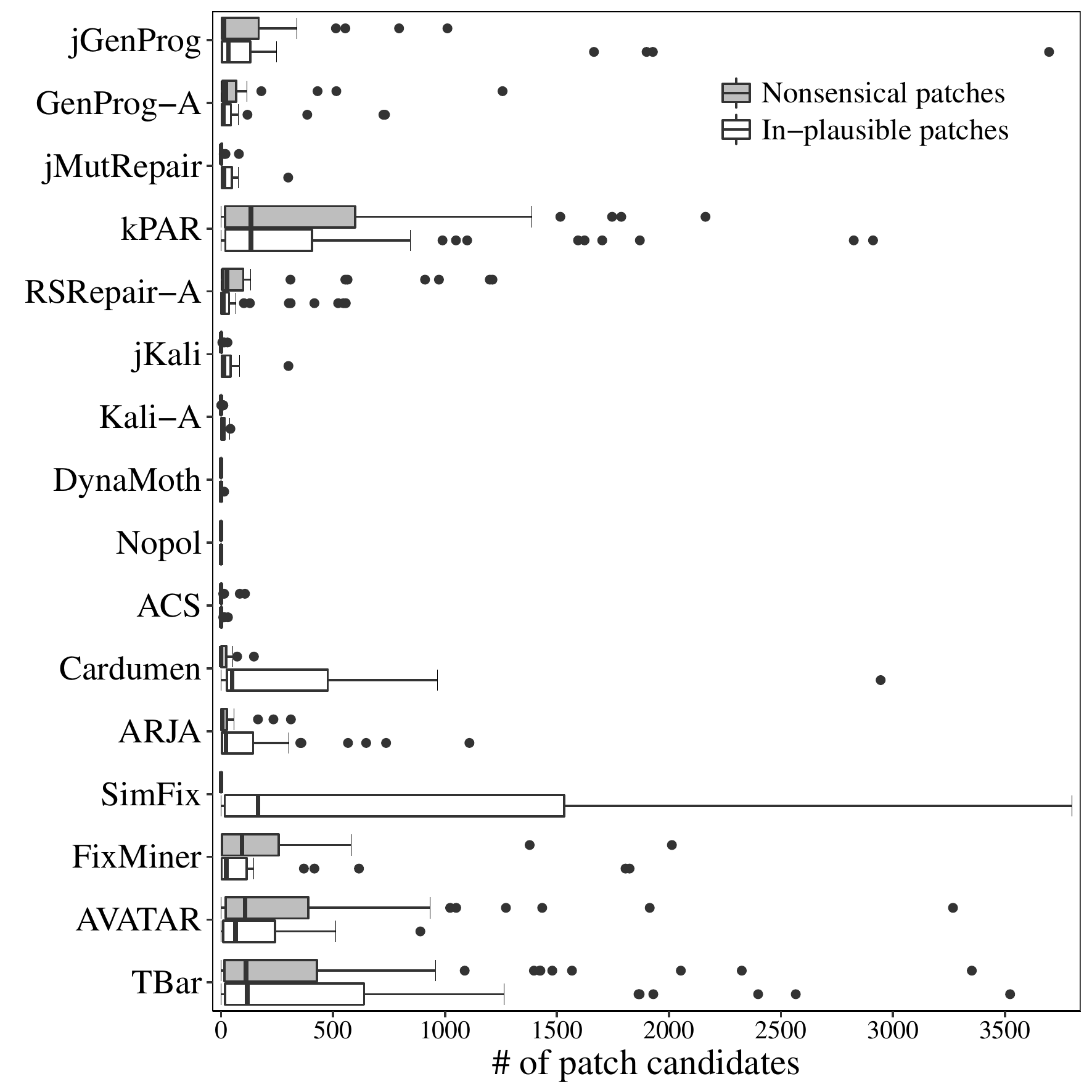}
    \caption{Distributions of $NPC_{nonsensical}$ and $NPC_{in-plausible}$ scores for each APR tools.}
    \label{fig:NFL-NPC2}
\end{figure}

$\bullet$ [{\em The state-of-the-art can avoid generating nonsensical patches.}]
Figure~\ref{fig:NFL-NPC2} illustrates the contribution of nonsensical and in-plausible patches to the $NPC$ scores. The distributions of nonsensical patches are interesting with respect to different claims in the literature. Indeed, to motivate their seminal work on template-based program repair, Kim et al.~\cite{kim2013automatic}, authors of the PAR tool, stated that pioneer genetic programming based repair tools had the limitation that it could generate nonsensical patches.
Our empirical assessment results back up this claim. However, our results also reveal that template-based repair tools (e.g., kPAR and TBar) have not fulfilled the claimed promise since they produce the largest numbers of nonsensical patches. This finding calls for a triaging strategy targeting nonsensical patches within the search space. In this regard, our experimental results highlight three tools (i.e., DynaMoth, Nopol, and SimFix), which do not generate any nonsensical patches.

Nopol uses an SMT solver to address the condition patch synthesis problem.
DynaMoth leverages the runtime context, collects variable and method calls to synthesize conditional expression patches.
SimFix heuristically searches similar code from the intersection of two search spaces: one is for donor code and the other one is for code change actions, to generate patches. A noteworthy result is that, while Nopol and DynaMoth overall generate few candidates, SimFix generates the largest number of patch candidates, none of which is ever found nonsensical. This finding suggests that code similarity has a large influence and can be useful for effectively triaging the repair search space.

Besides Nopol, Dynamoth, and SimFix, five repair tools (i.e, jMutRepair, jKali, Kali-A, Cardumen and ARJA) generate significantly more in-plausible patches than nonsensical ones.
jMutRepair, jKali and Kali-A are implemented with simple mutation operators that are unlikely to prevent the programs from compiling.  However, these mutation operations can lead to test failures.
ARJA's efficiency w.r.t. nonsensical patch generation is likely due to the combination of different search strategies that drive its genetic programming.

$\bullet$ [{\em The more templates an APR system considers, the more nonsensical and in-plausible patches it will generate.}]
TBar contains more fix templates than kPAR, FixMiner and AVATAR since it merges all literature templates. Therefore, each suspicious buggy location has a higher probability in TBar to be matched with more templates, leading to more patch candidates than other tools.
This finding highlights the importance of strategies for fix template matching and donor code searching to
 improve the repair efficiency of template-based repair tools.

$\bullet$ [{\em Specialized templates increase the efficiency of APR tools.}] Among the template-based repair tools, kPAR has the smallest number of templates. Indeed it includes 10 templates manually prepared by Kim et al.~\cite{kim2013automatic}, while AVATAR includes 11, TBar integrates 35 and FixMiner considers 28. Nevertheless, experimental results for NPC scores (cf. Figure~\ref{fig:NFL-NPC1}) and the dissection in non-sensical and in-plausible categories (cf. Figure~\ref{fig:NFL-NPC2}) reveal that kPAR is the least efficient. According to the authors' source code of the tools, these tools use the same search space traversal strategy and implementation. Therefore, the only difference being about the included templates, we can safely conclude that the nature of these templates is driving the efficiency performance.
AVATAR indeed focuses on templates obtained by curated datasets of fixes: all mined code changes are for static analysis violations which are systematically validated as actual fixes. FixMiner, on the other hand, augments its templates with relevant contextual information to ensure that they are applied on code locations that are syntactically similar to the locations where the templates where mined.

\begin{figure}[!h]
    \centering\vspace{-1mm}
    \includegraphics[width=\linewidth]{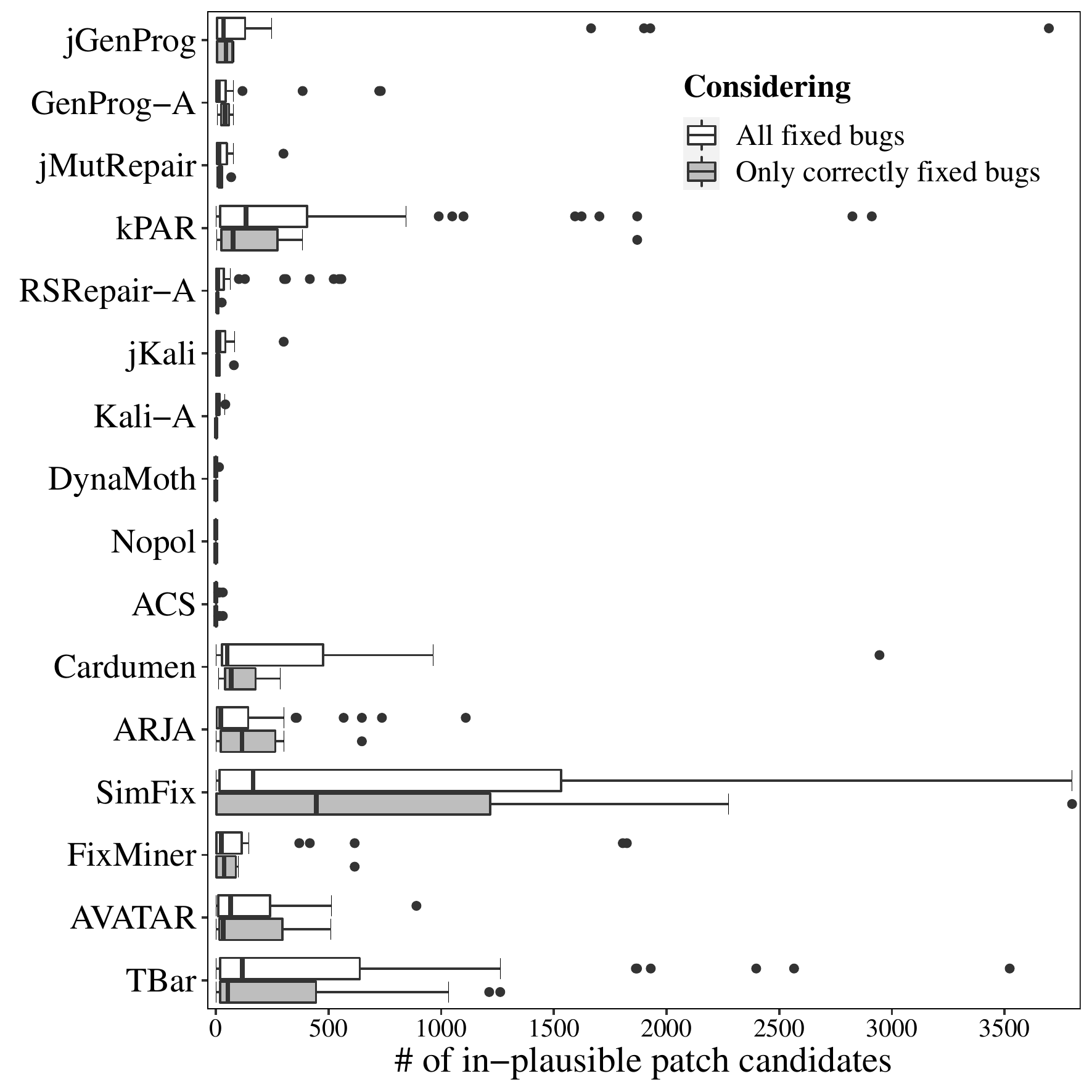}\vspace{-0.5mm}
    \caption{Number of in-plausible patch candidates generated before the first plausible patch.}
    \label{fig:NFL-NPC3}
\end{figure}

$\bullet$ [{\em Correct patches are sparse in the search space.}] Long et
al.~\cite{long2016analysis} presented an initial study which revealed that
correct patches can be considered as sparse in the search space and that
overfitting patches~\cite{qi2015analysis,xiong2018identifying,le2018overfitting,le2019reliability}
(i.e., only plausible but not correct) are vastly more abundant. We
extend their study to consider the cases of in-plausible patches that are
produced "before any plausible patch" (i.e., including if it is correct) vs. "before a correct patch" (i.e., only if the plausible is correct).
Figure~\ref{fig:NFL-NPC3} illustrates the distributions of $NPC_{in-plausible}$
scores for all fixed bugs and only correctly-fixed ones.
We observe that for tools such as TBar, AVATAR, FixMiner, and kPAR, the median of
$NPC_{in-plausible}$ scores for correctly-fixed bugs is lower than the median for all fixed bugs. This means that, when a correct patch can be found, the number of in-plausible patches that are generated before is fewer than when only a plausible patch can be found.
The situation is the converse for SimFix and ARJA.
Therefore, we note that for most tools, a correct patch is more
efficiently found when the search space is less noised (i.e., fewer in-plausible
patches).

\vspace{-1mm}
Table~\ref{tab:NFL_NPC} provides more detailed statistics to drive an in-depth
correlation study around efficiency and correctness. Based on the mean values,
except for ACS, ARJA, and AVATAR, APR tools tend to generate more patch candidates when considering all bugs than when considering only the correctly-fixed ones. This
tendency is much more apparent for {\em search-based} APR techniques such as
jGenProg~\cite{martinez2016astor},  GenProg-A~\cite{yuan2018arja},
SimFix~\cite{jiang2018shaping}, and RSRepair-A~\cite{yuan2018arja}. Although
TBar is a template-based approach, it has characteristics of search-based
tools since its search-space has been enlarged by incorporating {\bf any} fix templates in
the literature.

\begin{table}[!t]
	\center
	\scriptsize
	\caption{Upper whisker, median and mean values of $NPC$ ($NPC_{in-plausible}$) scores in Figures~\ref{fig:NFL-NPC1} and	~\ref{fig:NFL-NPC3}.}
	\label{tab:NFL_NPC}
	\resizebox{1\linewidth}{!}{
	\begin{threeparttable}
		\begin{tabular}{l|rr|rr|rr|r}
			\toprule
			\multirow{2}{*}{APR Tools} &\multicolumn{2}{c|}{Upper Whisker} & \multicolumn{2}{c|}{Median} & \multicolumn{2}{c|}{Mean} & \multirow{2}{*}{\makecell[c]{\#\\bugs}}\\\cline{2-7}
			& All & Correct & All & Correct  & All & Correct & \\
			\hline
			jGenProg   & 803 (247) & 191 (79)    & 50 (34) &$\uparrow$ 64 (45)& 670 (436) & 79 (41) & 5\\
			GenProg-A  & 235 (76)  & 139 (40)    & 34 (11) &$\uparrow$ 75 (41)  & 187 (81)  & 75 (40) & 2\\
			jMutRepair &  67 (77)  & 33 (27)     & 20 (14) &$\uparrow$ 28 (13) & 43 (36)  & 32 (27)   & 5\\
			kPAR       &2377 (844) & 992 (383)   &269 (134)&130 (68) & 879 (480) &600 (298)           & 10\\
			RSRepair-A &  208 (65) & 20 (18)    & 34 (10) &15 (6)  & 250 (81)  & 35 (10) & 4\\
			jKali      &  92 (83)  & 18 (16)     & 14 (13) & 7 (5)   & 35 (31)   & 21 (19)            & 6\\
			Kali-A     & 43 (38)   & 4 (3)       & 8 (7)   & 2 (1)   & 12 (10)   & 3 (2)              & 3\\
			Dynamoth   & 1 (0)     & 1 (0)       & 1 (0)   & 1 (0)   & 2 (1)     & 1 (0)              & 1\\
			Nopol      & 1 (0)     & 1  (0)      & 1 (0)   & 1 (0)   & 1 (0)     & 1 (0)              & 1\\
			ACS        & 15 (4)    & 15 (3)      & 2 (0)   & 2 (0)   & 15 (4)    &$\uparrow$ 17 (4)   & 17\\
			Cardumen   & 966 (965) & 286 (285)    & 87 (50) & 141 (68) & 479 (454) & 147 (122)            & 3\\
			ARJA       & 362 (302) & $\uparrow$ 686 (648)&38 (22)&$\uparrow$ 87 (83)&142 (117)&$\uparrow$ 211 (197)  & 6\\
			SimFix     &3801 (3800)&2274 (2273)  &164 (163)&$\uparrow$ 447 (446)&1168 (1167)&895 (894)           & 25\\
			FixMiner   & 546 (147) & 357 (99)    &111 (24) & 77 (24) & 754 (189) &656 (87)            & 12\\
			AVATAR     &1624 (512) &$\uparrow$ 2426 (511)&164 (65) &136 (33) &478 (145)&$\uparrow$ 530 (150)&19\\
			TBar       &2958 (1262)& 1806 (1031) &240 (118)&120 (53) & 818 (444) &620 (306)           & 24\\
			\bottomrule
		\end{tabular}
		{\footnotesize$^\ast$The upper whisker value is determined by 1.5 IQR (interquartile ranges) where IQR = 3rd Quartile - 1st Quartile, as defined in~\cite{frigge1989some}.
		``All'' denotes all fixed bugs, and ``Correct'' denotes correctly fixed bugs.
		The numbers outside the parentheses indicate the related $NPC$ score values and the numbers inside the parentheses indicate the related $NPC_{in-plausible}$ score values.
		$\uparrow$ implies that the $NPC$ and $NPC_{in-plausible}$ values of ``Correct'' are higher than those of ``All''.
		``\# bugs'' denotes the number of bugs correctly fixed by each repair tool.}
	\end{threeparttable}
	}
\end{table}

\vspace{-0.5mm}
The previous experimental data overall suggest that simply giving more time to the APR tool to repair a buggy program does not
guarantee to find correct patches.
On the contrary, it seems that when allowing less attempts, the correctness ratio is improved. We propose to simulate a simple strategy of threshold setting to investigate the impact on the correctness ratio (i.e., ratio of correctly-fixed bugs to plausibly-fixed bugs). We consider a scenario where the APR tool is halted when a certain number of in-plausible patches is checked.

\begin{table}[!h]
	\center
	\scriptsize
	\caption{CR after setting a $NPC_{in-plausible}$ threshold.}
	\label{tab:NFL_NPC_threshold}
	\resizebox{1\linewidth}{!}{
	\begin{threeparttable}
		\begin{tabular}{l|rrr||l|rrr}
			\toprule
			{\bf Tool} & {\bf TH$^\ast$} & \makecell[r]{{\bf\# fixed}\\{\bf bugs}} & {\bf CR(\%)} &
			{\bf Tool} & {\bf TH$^\ast$} & \makecell[r]{{\bf\# fixed}\\{\bf bugs}} & {\bf CR(\%)} \\
			\midrule
			jGenProg   & 80 & 5 (14) & +10.7& Nopol      & 0  & 1 (31) & 0    \\
			GenProg-A  & 80 & 2 (25) & +1.3 & ACS        & 32 &17 (22) & 0    \\
			jMutRepair & 70 & 5 (20) & +2.3 & Cardumen   & 70 & 2 (7)  & +11.9\\
			kPAR       & 300& 8 (42) & +3.1 & ARJA       & 650& 6 (56) & +0.4 \\
			RSRepair-A & 26 & 4 (27) & +4.8 & SimFix     &3800&24 (61) & +4.0 \\
			jKali      & 80 & 6 (22) & +3.3 & FixMiner   & 100&11 (23) & +11.4 \\
			Kali-A     & 3  & 3 (26) & +6.9 & AVATAR     & 511&19 (55) & +0.2 \\
			Dynamoth   & 0  & 1 (21) & +0.2 & TBar       &1230&24 (66) & +5.6 \\
			\bottomrule
		\end{tabular}
		{\footnotesize$^\ast$The threshold (TH) for each repair tool is set with its upper-bound
		$NPC_{in-plausible}$ score
		shown in Figure~\ref{fig:NFL-NPC3}.}
	\end{threeparttable}
	}
\end{table}

Table~\ref{tab:NFL_NPC_threshold} presents the results on how correctness ratio is influenced when we set a threshold on the number of in-plausible patches: basically, we propose to stop the repair attempts by a given tool if a certain number of generated patches turned out to be in-plausible (i.e., do not pass the test cases).
We observe that the ratio of generated plausible patches to be correct is increased at varying degrees for 14 (out of 16) repair tools. Nopol and ACS do not show any improvement: initially, they produce few in-plausible patches. It should be noted that this result should be put in perspective as when discussing precision and recall: threshold setting, while useful to increase correctness ratio, may also lead to an overall reduction of the number of bugs that are correctly fixed.

\notez{Overall our systematic study of patch generation efficiency reveals that (1) efficiency is not yet a widely-valued performance target; (2) state-of-the-art can avoid generating nonsensical patches; (3) the more templates an APR system considers, the more nonsensical and in-plausible patches it will generate; (4) specialized templates increase  APR tool efficiency; and (5) correct patches are sparse in the search space.}

\subsection{RQ3: Impact of Fault Localization Noise}

A recent study by Liu et al.~\cite{liu2019you} has reported empirical results
suggesting that fault localization results can adversely affect the performance of the repair. The authors experimented on a single tool, kPAR, and focused on repairability (i.e., how many bugs are not fixed due to localization errors).
Our study already takes steps to avoid the bias of presenting various experimental results with APR tools which use different fault localization inputs. Thus, we have put an effort to harmonize all fault localization configurations for the 16 APR tools under study (cf. Section~\ref{sec:config}).

To evaluate the impact of fault localization noise for different tools, we propose to compare results obtained so far with our standard spectrum-based fault localization (GZoltar+Ochiai) against experimental results where the APR systems are directly given the ground-truth fix locations.
We compare the results both in terms of repairability and repair efficiency.


\vspace{-1.5mm}
\subsubsection{Impact of fault localization noise on repairability} First, we measure the impact on repairability, where we estimate for each repair tool {\bf how many bugs can be fixed by each APR system if it is precisely pointed to the ground-truth fix locations?}
Table~\ref{tab:PFL_REPAIR} illustrates the details on the impact of repairability.
Except for Cardumen, we observe that in general the correctness ratio improves (by up to 30 percentage points) if the fix locations are provided. It suggests that false-positive bug locations, hence fault localization noise, has an impact on the likelihood to generate correct patches.
There are however anecdotical cases that are noteworthy:

$\bullet$ [{\em Ground truth incompleteness.}] Although our configuration of fault localization did not yield the developer-provided fix position for bug {\tt Lang}-35, ACS patch generation eventually produced a correct patch for this bug. This patch, which targets a different code location, was found semantically-similar to the developer-provided patch following rule R2 (cf. Section~\ref{sec:validation}). This finding reminds us that the benchmark that is used is not a complete ground-truth, neither for repair-oriented fault localization nor for patch generation.

\begin{table}[!ht]
	\centering
	\scriptsize
	\caption{Impact$^\dagger$ on repairability$^\ast$ when ground-truth fix locations are directly given to the APR system.}
	\label{tab:PFL_REPAIR}
	\resizebox{1\linewidth}{!}{
	\begin{threeparttable}
		\begin{tabular}{lcccccccc}
			\toprule
			{\bf APR Tool} & {\bf C} & {\bf Cl} & {\bf L} & {\bf M} & {\bf Mc} & {\bf T} & {\bf Total} & {\bf CR (\%)}\\
			\hline
			jGenProg   & +1 (-3) & -1 (0)  & 0 (-2)  & +1 (+1) & 0 (0) & 0 (0) & +1 (-4)& +12.5\\
			GenProg-A  & 0 (-2)  & +3 (+1) & +1 (+2) & +2 (-2) & 0 (0) & 0 (0) & +6 (-1)& +20.9\\
			jMutRepair & 0 (-3)  & 0 (-1)  & 0 (-2)  & 0 (-5)  & 0 (0) & 0 (0) & 0 (-11)& +22.8\\
			kPAR       & +3 (-5) &+11 (+11)& +3 (-5) & +3 (-6) & 0 (0) &+3 (+4)&+23 (0) & +36.5\\
			RSRepair-A & 0 (-2)  & 0 (-6)  & +1 (+1) & +4 (0)  & 0 (0) & 0 (0) &+5 (-7) & +16.5\\
			jKali      & 0 (-3)  &-1 (-6)  & -2 (-4) & -1 (-4) & 0 (0) & 0 (0) &-4 (-17)& +1\\
			Kali-A     & 0 (-5)  & +2 (-18)& 0 (+3)  & 0 (-2)  & 0 (-1)& 0 (0) &+2 (-23)& +7.3\\
			DynaMoth   & 0 (-5)  & N/A     & +2 (+2) & 0 (-5)  & 0 (0) & 0 (-1)& +2 (-9)& +18.6\\
			Nopol      & 0 (-5)  & N/A     & 0 (-3)  & +1 (-13)& 0 (0) & 0 (-1)& +1 (-22)& +19\\
			ACS        & 0 (0)   & 0 (0)   & -1 (-1) & 0 (0)   & 0 (0) & 0 (0) &-1 (-1) & +3.5\\
			Cardumen   & -1 (+2) & 0 (-2)  & 0 (+1)  & 0 (+3)  & 0 (0) & 0 (0) & -1 (+4)& -4.2\\
			ARJA       & 0 (-8)  & +2 (-13)& +1 (+2) & +2 (-2) & 0 (-1)& 0 (0) &5 (-22)& +20.2\\
			SimFix     & +1 (-4) & 0 (-2)  & 0 (-10) &+2 (-4)  & 0 (0) &+1 (+2) &+4 (-18)& +21.2	\\
			FixMiner   & +2 (-5) & +6 (+13)&	 +4 (+7) & +5 (+10)&+2 (+2)&+3 (+3)&+22 (+30)& +14.6\\
			AVATAR     & +1 (-4) & +2 (-2) & +1 (-2) & +3 (-4) &+2 (+2)&+2 (+3)&+11 (-7)& +26.7\\
			TBar       & +3 (-3) &+12 (+12)&+4 (-3)  & +5 (-1) &+3 (+3)&+3 (+5)&+30 (+13)& +32.7\\
			\bottomrule
		\end{tabular}	
		{\footnotesize$^\dagger$This table shows variations of repairability w.r.t. results of our generic configuration of fault localization provided in Table~\ref{tab:NFL}. $^\ast${\em +x(-y) means that, if given exact fix locations, the tool can correctly fix x more bugs, but plausibly fixes y less bugs}\\
		}
	\end{threeparttable}
	}
\end{table}

\vspace{-1.5mm}
$\bullet$ [{\em Fix location is different from bug location.}] We observe that jKali now fails to correctly fix respectively 2 when it is given the developer-provided fix locations. This finding suggests that the repair tool is rather misled, in the cases of specific bugs, when it is given the right bug positions. Instead, some sibling positions are better inputs to drive correct fixing. However, data in Table~\ref{tab:PFL_REPAIR} show fault localization has different impacts on performance for plausible fixing than for correct fixing.

\begin{figure}[!t]
    \centering
    \includegraphics[width=1\linewidth]{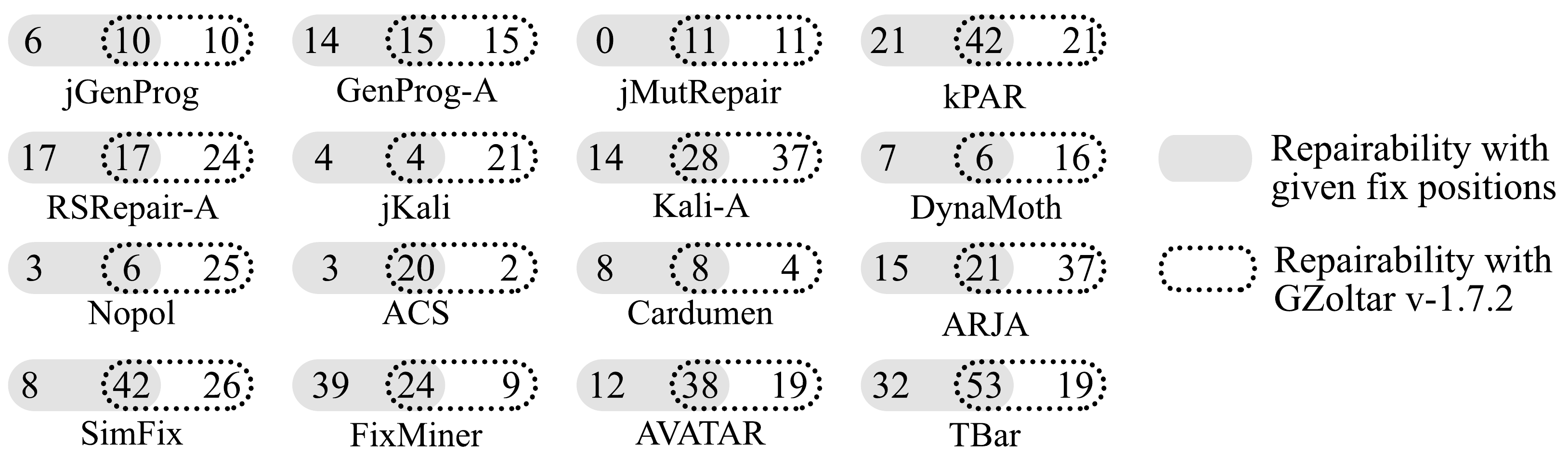}
    \vspace{-4mm}
    \caption{Overlap and difference between normal fault localization and given fix positions for repair tools.}
    \vspace{-1mm}
\label{fig:overlap_PF}
\end{figure}

Furthermore, based on results of overlapping in repairability (in terms of plausible patches) performance as depicted in Figure~\ref{fig:overlap_PF}, we note that many bugs are only fixed (plausibly) when the fault localization does not precisely point to the fix locations. This is a surprising but interesting finding to be investigated by APR-targeted fault localization research.

$\bullet$ [{\em Mockito bugs are not repairable.}]
Another immediate observation that we make from the experimental results in Table~\ref{tab:PFL_REPAIR} is that bugs from the {\tt Mockito} project are not easy to fix. According to reported results in Table~\ref{tab:PFL_REPAIR}, only three tools (i.e., FixMiner, AVATAR, and TBar) are able to fix {\tt Mockito} bugs even if ground-truth fix locations are provided. We carefully proceed to investigate the possible reasons for this situation: 13 {\tt Mockito} bugs (i.e., bug IDs 1-10 and 18-20) are associated to program code that cannot be compiled under JDK 7 (which is the JDK that is mentioned in the requirements of Defects4J). Our results further confirm a recent study~\cite{wang2019attention} by Wang et al., who reported that the state-of-the-art SimFix and CapGen are not able to fix any {\tt Mockito} bugs even when provided with ground-truth fix locations. Our study enlarges the scope of their studies. In the end, our systematic assessment results for all bugs better sheds light on a common phenomenon in the literature where {\tt Mockito} project bugs are not considered when reporting repair performance. These results call for modular configuration of execution environment as well as for better integration of advances in fault localization to support APR systems.
Besides Mockito bugs, many bugs in other projects cannot be fixed since they are not precisely localized. Overall, consider again Figure~\ref{fig:overlap_PF}. For all tools (except jMutRepair), we observe that some bugs are fixed only when the actual fix locations are directly given to the system.



\subsubsection{Impact of fault localization noise on repair efficiency}
\begin{figure}[!h]
    \centering
    \includegraphics[width=\linewidth]{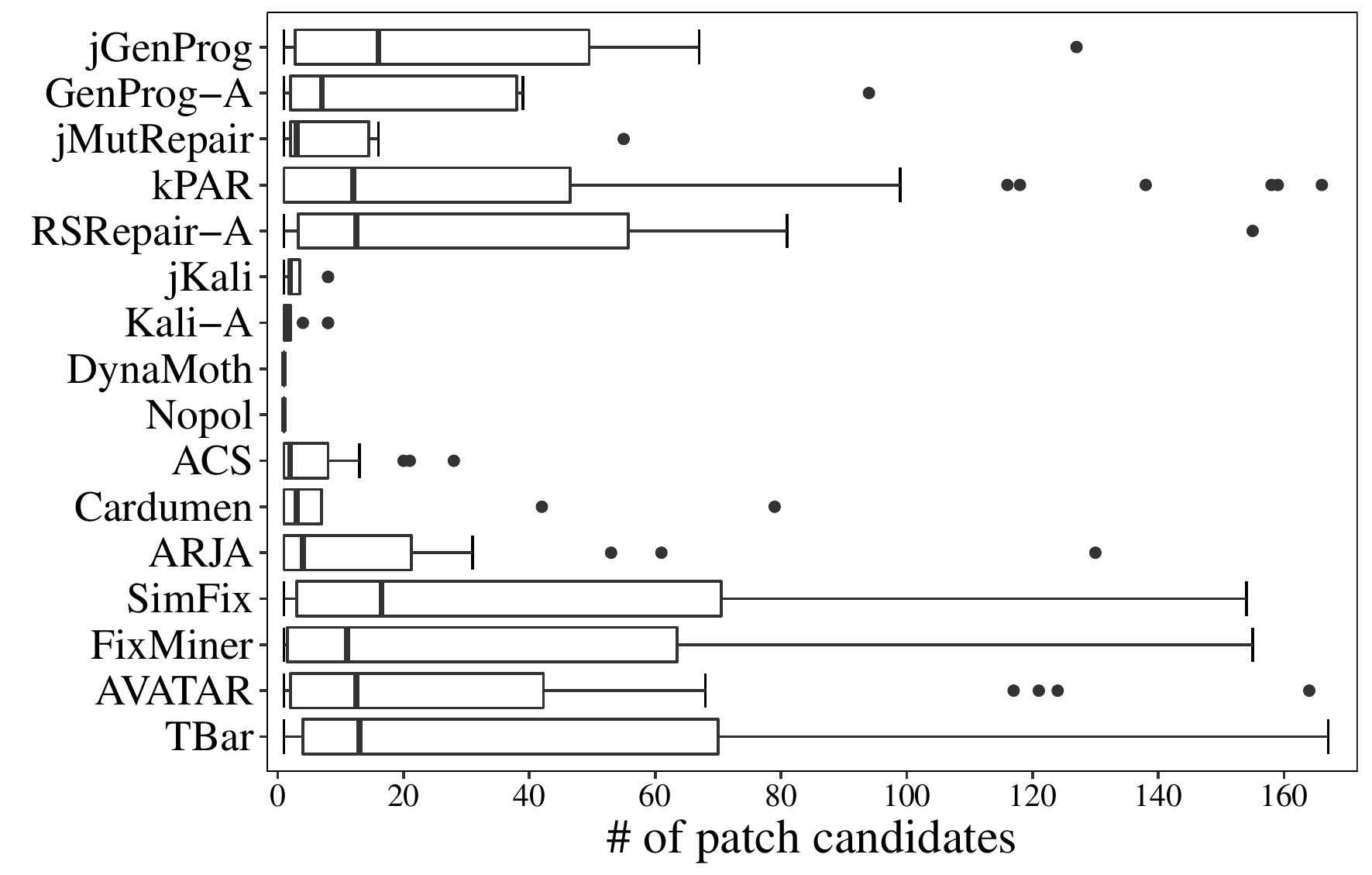}
    \vspace{-4mm}
    \caption{NPC score distribution of each tool given fix positions.}
    \vspace{-2mm}
    \label{fig:PFL_NPC}
\end{figure}

We investigate the $NPC$ scores, i.e., the number of patch candidates that are generated by the different APR systems when they are pointed to the developer-provided fix locations.
Figure~\ref{fig:PFL_NPC} shows the corresponding distribution of $NPC$ scores for each repair tool.

$\bullet$ [{\em Template-based program repair tools are highly sensitive to fault localization noise.}] We observe from Figure~\ref{fig:PFL_NPC} that, except for DynaMoth, Nopol, and ACS, the remaining 13 repair tools have significantly smaller distribution ranges of $NPC$ scores than the
 distribution ranges when the APR system was run under our generic fault localization configuration (cf. Figure~\ref{fig:NFL-NPC1}).
A straightforward explanation is that, under a typical fault localization configuration, a repair tool will attempt to generate patch candidates for each suspicious statement that is ranked by the fault localization.
When the fault localization is noisy (i.e., top suspicious statement(s) are false positives), more in-plausible and even non-sensical patches might be generated.
In particular, for repair tools that are based on pattern matching and code similarity (i.e., SimFix, and the template-based repair tools) the gap of repair efficiency has reduced substantially by an order of magnitude when correct fix locations are given to the tool. For example, the median $NPC$ score of SimFix is around 200 when using our generic configuration of fault localization, but is around 20 when using directly correct fix locations. Such tools are thus more sensitive to fault localization noise than other tools. In conclusion, we confirm the finding of the study of Liu et al.~\cite{liu2019you}. However, we delimitate its validity to template-based repair tools. Other tools, e.g., constraint-based repair tools such as ACS or Nopol, which use specific techniques to triage the search space do not present any increase in repair efficiency when pointed to the fix locations. This finding suggests that they have limited sensitivity to fault localization noise.

\notez{Fault localization is an important step in a repair pipeline. Its false positives, however, have a significant impact on both repairability and repair efficiency. In particular, we found that accurately localizing the bug can reduce the number of generated patches by an order of magnitude, thus drastically enhancing efficiency. From the perspective of repairability, better fault localization will increase the probability to generate correct patches (i.e., the correctness ratio).}

\vspace{-2mm}
\section{Threats to Validity}
\label{sec:dis}

%
%
%
%
{\em External validity.} Our study considers only the Defects4J benchmark and only java repair tools. All findings might thus be valid only for this configuration. Nevertheless, this threat is mitigated by the fact that we use a large set of repair tools and a renowned defect benchmark to study a performance criterion that was largely ignored in the literature.

{\em Internal validity.} Our implementation of fault localization as well as the manual assessment of patch correctness may threaten the validity of some of our conclusions. We mitigate this threat by reusing common fault localization components from the repair literature as well as by enumerating and sharing the rules for defining patch correctness. Two authors were in charge of assessing the correctness and they cross-reviewed each other's decisions. In case of conflict other authors were called to create a consensus.

{\em Construct validity.} By construct, to limit resource exhaustion, we added a threshold on the number of patches to validate. However, this threshold may penalize some tools. We mitigate this threat by carefully selecting a threshold based on empirical results on PraPR, a recent related work which mutates directly bytecode, allowing it to generate many more patches (since no compilation is needed).

\section{Related Work}
\label{sec:relatedWork}

\paragraph{\bf Performance Evaluation}
Initially, evaluation of test-based program repair was focused on counting the number of bugs fixed by a repair tool out of all bugs in a benchmark~\cite{weimer2009automatically,le2012genprog,kim2013automatic,le2016history}. 
However, valid patches are sometimes incorrect as they overfit on incomplete test suites~\cite{qi2015analysis}, and might cause issues during maintainance~\cite{fry2012human,smith2015cure}. Thus, plausibility and correctness became widely accepted to define metrics for assessing repairability of repair tools~\cite{xiong2017precise,wen2018context,chen2017contract,hua2018towards,saha2017elixir,saha2019harnessing,ghanbari2019practical,liu2018lsrepair,liu2019tbar,liu2019avatar,liu2019you,koyuncu2018fixminer}.
In this study, we also follow the metric to revisit the repairability of repair tools. Nevertheless, we differ from studies in the literature by ensuring that APR tools use the same controlled configuration for fault localization.

\paragraph{\bf Repair Efficiency}
Along with the performance evaluation, serval studies simply reported the repair efficiency in terms of CPU time consumption of fixing bugs~\cite{weimer2009automatically,xiong2017precise,wen2018context,hua2018towards,ghanbari2019practical}.
However, it could be biased to assess the efficiency with time cost for various reasons (cf. Section~\ref{sec:bg}).
Instead, we leverage the number of patch candidates generated by repair tools to measure the repair efficiency, which should be intrinsic to the repair approaches. Ghanbari et al.~\cite{ghanbari2019practical} provided information on the number of patch candidates generated by PraPR. This information, however, could not be put into perspective against other tools. Our study fills this gap.

\paragraph{\bf Empirical Study}
To boost the development of program repair, various empirical studies have been conducted in this direction.
Le Goues et al.~\cite{le2012systematic} re-assessed GenProg on real bugs, while several studies on overfitting followed~\cite{qi2013using,qi2015analysis,le2018overfitting,xiong2018identifying,le2019reliability,wang2019different}.
Yang et al.~\cite{yang2017better} explored better test cases for better program repair. 
Yi et al.~\cite{yi2018correlation} empirically investigated the effectiveness of test-suite metrics in controlling the repairing reliability of GenProg. 
Motwani et al.~\cite{motwani2018automated} investigated to what extent important bugs can be fixed by 9 APR tools.
Liu et al.~\cite{liu2019you} investigated the FL bias in benchmarking APR tools with only one APR tool.
Durieux et al.~\cite{durieux2019empirical} conducted a large-scale empirical study for Java APR tools to investigate their repairability on different benchmarks.
Empirical studies for APR tools have been studied from different scenarios in the literature, but these studies mainly focus on the traditional APR tools and the latest state-of-the-art tools (e.g., ACS~\cite{xiong2017precise}, SimFix~\cite{jiang2018shaping} and TBar~\cite{liu2019tbar}) have not been studied systematically. Our study fills this gap by looking back at 10 years of test-based program repair research and focusing on the under-valued performance criterion that is efficiency.

\section{Conclusion}
\label{sec:conc}

This paper reports on a large-scale study on the efficiency of test suite based program repair. Efficiency is defined based on the number of patch candidates that are generated before a repair system can hit a valid patch. Our study comprehensively runs 16 repair systems from the literature under identical configuration of fault localization. Our experiments explore repairability (i.e., repair effectiveness), repair efficiency as well as the impact of fault localization on both performance criteria. Beyond the statistical data, we call on the community to {\bf invest in strategies for making repair efficient} in order to facilitate adoption in a software industry where computing resources are managed sometimes with parsimony.
 
\noindent
{\bf Artefacts:} All data and tool support for replication are available at
	\url{https://github.com/SerVal-DTF/APR-Efficiency.git}

\begin{acks}
This work is supported by the Fonds National de la Recherche (FNR), Luxembourg, through RECOMMEND 15/IS/10449467 and FIXPATTERN C15/IS/9964569, and is supported through the National Natural Science Foundation of China No. 61672529.
\end{acks}

\balance
\bibliographystyle{ACM-Reference-Format}
\bibliography{bib/references}


\begin{thebibliography}{67}


\ifx \showCODEN    \undefined \def \showCODEN     #1{\unskip}     \fi
\ifx \showDOI      \undefined \def \showDOI       #1{#1}\fi
\ifx \showISBNx    \undefined \def \showISBNx     #1{\unskip}     \fi
\ifx \showISBNxiii \undefined \def \showISBNxiii  #1{\unskip}     \fi
\ifx \showISSN     \undefined \def \showISSN      #1{\unskip}     \fi
\ifx \showLCCN     \undefined \def \showLCCN      #1{\unskip}     \fi
\ifx \shownote     \undefined \def \shownote      #1{#1}          \fi
\ifx \showarticletitle \undefined \def \showarticletitle #1{#1}   \fi
\ifx \showURL      \undefined \def \showURL       {\relax}        \fi
\providecommand\bibfield[2]{#2}
\providecommand\bibinfo[2]{#2}
\providecommand\natexlab[1]{#1}
\providecommand\showeprint[2][]{arXiv:#2}

\bibitem[\protect\citeauthoryear{Abreu, Zoeteweij, and Van~Gemund}{Abreu
  et~al\mbox{.}}{2007}]%
        {abreu2007accuracy}
\bibfield{author}{\bibinfo{person}{Rui Abreu}, \bibinfo{person}{Peter
  Zoeteweij}, {and} \bibinfo{person}{Arjan~JC Van~Gemund}.}
  \bibinfo{year}{2007}\natexlab{}.
\newblock \showarticletitle{On the accuracy of spectrum-based fault
  localization}. In \bibinfo{booktitle}{\emph{Testing: Academic and Industrial
  Conference Practice and Research Techniques-MUTATION}}. IEEE,
  \bibinfo{pages}{89--98}.
\newblock
\urldef\tempurl%
\url{https://doi.org/10.1109/TAIC.PART.2007.13}
\showDOI{\tempurl}


\bibitem[\protect\citeauthoryear{Allamanis, Barr, Devanbu, and
  Sutton}{Allamanis et~al\mbox{.}}{2018}]%
        {allamanis2018survey}
\bibfield{author}{\bibinfo{person}{Miltiadis Allamanis},
  \bibinfo{person}{Earl~T. Barr}, \bibinfo{person}{Premkumar~T. Devanbu}, {and}
  \bibinfo{person}{Charles~A. Sutton}.} \bibinfo{year}{2018}\natexlab{}.
\newblock \showarticletitle{A Survey of Machine Learning for Big Code and
  Naturalness}.
\newblock \bibinfo{journal}{\emph{Comput. Surveys}} \bibinfo{volume}{51},
  \bibinfo{number}{4} (\bibinfo{year}{2018}), \bibinfo{pages}{81:1--81:37}.
\newblock
\urldef\tempurl%
\url{https://doi.org/10.1145/3212695}
\showDOI{\tempurl}


\bibitem[\protect\citeauthoryear{Chen, Pei, and Furia}{Chen
  et~al\mbox{.}}{2017}]%
        {chen2017contract}
\bibfield{author}{\bibinfo{person}{Liushan Chen}, \bibinfo{person}{Yu Pei},
  {and} \bibinfo{person}{Carlo~A Furia}.} \bibinfo{year}{2017}\natexlab{}.
\newblock \showarticletitle{Contract-based program repair without the
  contracts}. In \bibinfo{booktitle}{\emph{Proceedings of the 32nd IEEE/ACM
  International Conference on Automated Software Engineering}}. IEEE,
  \bibinfo{pages}{637--647}.
\newblock
\urldef\tempurl%
\url{https://doi.org/10.1109/ASE.2017.8115674}
\showDOI{\tempurl}


\bibitem[\protect\citeauthoryear{Debroy and Wong}{Debroy and Wong}{2010}]%
        {debroy2010using}
\bibfield{author}{\bibinfo{person}{Vidroha Debroy} {and}
  \bibinfo{person}{W~Eric Wong}.} \bibinfo{year}{2010}\natexlab{}.
\newblock \showarticletitle{Using mutation to automatically suggest fixes for
  faulty programs}. In \bibinfo{booktitle}{\emph{Proceedings of the 3rd
  International Conference on Software Testing, Verification and Validation}}.
  IEEE, \bibinfo{pages}{65--74}.
\newblock
\urldef\tempurl%
\url{https://doi.org/10.1109/ICST.2010.66}
\showDOI{\tempurl}


\bibitem[\protect\citeauthoryear{Durieux, Cornu, Seinturier, and
  Monperrus}{Durieux et~al\mbox{.}}{2017}]%
        {durieux2017dynamic}
\bibfield{author}{\bibinfo{person}{Thomas Durieux}, \bibinfo{person}{Benoit
  Cornu}, \bibinfo{person}{Lionel Seinturier}, {and} \bibinfo{person}{Martin
  Monperrus}.} \bibinfo{year}{2017}\natexlab{}.
\newblock \showarticletitle{Dynamic patch generation for null pointer
  exceptions using metaprogramming}. In \bibinfo{booktitle}{\emph{Proceedings
  of the 24th International Conference on Software Analysis, Evolution and
  Reengineering}}. IEEE, \bibinfo{pages}{349--358}.
\newblock
\urldef\tempurl%
\url{https://doi.org/10.1109/SANER.2017.7884635}
\showDOI{\tempurl}


\bibitem[\protect\citeauthoryear{Durieux, Madeiral, Martinez, and
  Abreu}{Durieux et~al\mbox{.}}{2019}]%
        {durieux2019empirical}
\bibfield{author}{\bibinfo{person}{Thomas Durieux}, \bibinfo{person}{Fernanda
  Madeiral}, \bibinfo{person}{Matias Martinez}, {and} \bibinfo{person}{Rui
  Abreu}.} \bibinfo{year}{2019}\natexlab{}.
\newblock \showarticletitle{Empirical Review of Java Program Repair Tools: A
  Large-Scale Experiment on 2,141 Bugs and 23,551 Repair Attempts}. In
  \bibinfo{booktitle}{\emph{Proceedings of the 27th ACM Joint Meeting on
  European Software Engineering Conference and Symposium on the Foundations of
  Software Engineering}}. \bibinfo{publisher}{ACM}, \bibinfo{pages}{302--313}.
\newblock
\urldef\tempurl%
\url{https://doi.org/10.1145/3338906.3338911}
\showDOI{\tempurl}


\bibitem[\protect\citeauthoryear{Durieux and Monperrus}{Durieux and
  Monperrus}{2016}]%
        {durieux2016dynamoth}
\bibfield{author}{\bibinfo{person}{Thomas Durieux} {and}
  \bibinfo{person}{Martin Monperrus}.} \bibinfo{year}{2016}\natexlab{}.
\newblock \showarticletitle{{DynaMoth:} dynamic code synthesis for automatic
  program repair}. In \bibinfo{booktitle}{\emph{Proceedings of the 11th
  International Workshop in Automation of Software Test}}. ACM,
  \bibinfo{pages}{85--91}.
\newblock
\urldef\tempurl%
\url{https://doi.org/10.1145/2896921.2896931}
\showDOI{\tempurl}


\bibitem[\protect\citeauthoryear{Frigge, Hoaglin, and Iglewicz}{Frigge
  et~al\mbox{.}}{1989}]%
        {frigge1989some}
\bibfield{author}{\bibinfo{person}{Michael Frigge}, \bibinfo{person}{David~C
  Hoaglin}, {and} \bibinfo{person}{Boris Iglewicz}.}
  \bibinfo{year}{1989}\natexlab{}.
\newblock \showarticletitle{Some implementations of the boxplot}.
\newblock \bibinfo{journal}{\emph{The American Statistician}}
  \bibinfo{volume}{43}, \bibinfo{number}{1} (\bibinfo{year}{1989}),
  \bibinfo{pages}{50--54}.
\newblock


\bibitem[\protect\citeauthoryear{Fry, Landau, and Weimer}{Fry
  et~al\mbox{.}}{2012}]%
        {fry2012human}
\bibfield{author}{\bibinfo{person}{Zachary~P. Fry}, \bibinfo{person}{Bryan
  Landau}, {and} \bibinfo{person}{Westley Weimer}.}
  \bibinfo{year}{2012}\natexlab{}.
\newblock \showarticletitle{A human study of patch maintainability}. In
  \bibinfo{booktitle}{\emph{Proceedings of the 21st International Symposium on
  Software Testing and Analysis}}. \bibinfo{publisher}{{ACM}},
  \bibinfo{pages}{177--187}.
\newblock
\urldef\tempurl%
\url{https://doi.org/10.1145/2338965.2336775}
\showDOI{\tempurl}


\bibitem[\protect\citeauthoryear{Gazzola, Micucci, and Mariani}{Gazzola
  et~al\mbox{.}}{2019}]%
        {gazzola2019automatic}
\bibfield{author}{\bibinfo{person}{Luca Gazzola}, \bibinfo{person}{Daniela
  Micucci}, {and} \bibinfo{person}{Leonardo Mariani}.}
  \bibinfo{year}{2019}\natexlab{}.
\newblock \showarticletitle{Automatic software repair: A survey}.
\newblock \bibinfo{journal}{\emph{IEEE Transactions on Software Engineering}}
  \bibinfo{volume}{45}, \bibinfo{number}{1} (\bibinfo{year}{2019}),
  \bibinfo{pages}{34--67}.
\newblock
\urldef\tempurl%
\url{https://doi.org/10.1109/TSE.2017.2755013}
\showDOI{\tempurl}


\bibitem[\protect\citeauthoryear{Ghanbari, Benton, and Zhang}{Ghanbari
  et~al\mbox{.}}{2019}]%
        {ghanbari2019practical}
\bibfield{author}{\bibinfo{person}{Ali Ghanbari}, \bibinfo{person}{Samuel
  Benton}, {and} \bibinfo{person}{Lingming Zhang}.}
  \bibinfo{year}{2019}\natexlab{}.
\newblock \showarticletitle{Practical program repair via bytecode mutation}. In
  \bibinfo{booktitle}{\emph{Proceedings of the 28th ACM SIGSOFT International
  Symposium on Software Testing and Analysis}}. ACM, \bibinfo{pages}{19--30}.
\newblock
\urldef\tempurl%
\url{https://doi.org/10.1145/3293882.3330559}
\showDOI{\tempurl}


\bibitem[\protect\citeauthoryear{Gupta, Pal, Kanade, and Shevade}{Gupta
  et~al\mbox{.}}{2017}]%
        {gupta2017deepfix}
\bibfield{author}{\bibinfo{person}{Rahul Gupta}, \bibinfo{person}{Soham Pal},
  \bibinfo{person}{Aditya Kanade}, {and} \bibinfo{person}{Shirish Shevade}.}
  \bibinfo{year}{2017}\natexlab{}.
\newblock \showarticletitle{{DeepFix:} Fixing common {C} language errors by
  deep learning}. In \bibinfo{booktitle}{\emph{Proceedings of the 31st AAAI
  Conference on Artificial Intelligence}}. \bibinfo{publisher}{{AAAI}},
  \bibinfo{pages}{1345--1351}.
\newblock


\bibitem[\protect\citeauthoryear{Hua, Zhang, Wang, and Khurshid}{Hua
  et~al\mbox{.}}{2018}]%
        {hua2018towards}
\bibfield{author}{\bibinfo{person}{Jinru Hua}, \bibinfo{person}{Mengshi Zhang},
  \bibinfo{person}{Kaiyuan Wang}, {and} \bibinfo{person}{Sarfraz Khurshid}.}
  \bibinfo{year}{2018}\natexlab{}.
\newblock \showarticletitle{Towards practical program repair with on-demand
  candidate generation}. In \bibinfo{booktitle}{\emph{Proceedings of the 40th
  International Conference on Software Engineering}}. ACM,
  \bibinfo{pages}{12--23}.
\newblock
\urldef\tempurl%
\url{https://doi.org/10.1145/3180155.3180245}
\showDOI{\tempurl}


\bibitem[\protect\citeauthoryear{Jiang, Xiong, Zhang, Gao, and Chen}{Jiang
  et~al\mbox{.}}{2018}]%
        {jiang2018shaping}
\bibfield{author}{\bibinfo{person}{Jiajun Jiang}, \bibinfo{person}{Yingfei
  Xiong}, \bibinfo{person}{Hongyu Zhang}, \bibinfo{person}{Qing Gao}, {and}
  \bibinfo{person}{Xiangqun Chen}.} \bibinfo{year}{2018}\natexlab{}.
\newblock \showarticletitle{Shaping program repair space with existing patches
  and similar code}. In \bibinfo{booktitle}{\emph{Proceedings of the 27th ACM
  SIGSOFT International Symposium on Software Testing and Analysis}}. ACM,
  \bibinfo{pages}{298--309}.
\newblock
\urldef\tempurl%
\url{https://doi.org/10.1145/3213846.3213871}
\showDOI{\tempurl}


\bibitem[\protect\citeauthoryear{Just, Jalali, and Ernst}{Just
  et~al\mbox{.}}{2014}]%
        {just2014defects4j}
\bibfield{author}{\bibinfo{person}{Ren{\'e} Just}, \bibinfo{person}{Darioush
  Jalali}, {and} \bibinfo{person}{Michael~D Ernst}.}
  \bibinfo{year}{2014}\natexlab{}.
\newblock \showarticletitle{{Defects4J}: A database of existing faults to
  enable controlled testing studies for Java programs}. In
  \bibinfo{booktitle}{\emph{Proceedings of the 23rd International Symposium on
  Software Testing and Analysis}}. ACM, \bibinfo{pages}{437--440}.
\newblock
\urldef\tempurl%
\url{https://doi.org/10.1145/2610384.2628055}
\showDOI{\tempurl}


\bibitem[\protect\citeauthoryear{Kim, Nam, Song, and Kim}{Kim
  et~al\mbox{.}}{2013}]%
        {kim2013automatic}
\bibfield{author}{\bibinfo{person}{Dongsun Kim}, \bibinfo{person}{Jaechang
  Nam}, \bibinfo{person}{Jaewoo Song}, {and} \bibinfo{person}{Sunghun Kim}.}
  \bibinfo{year}{2013}\natexlab{}.
\newblock \showarticletitle{Automatic patch generation learned from
  human-written patches}. In \bibinfo{booktitle}{\emph{Proceedings of the 35th
  International Conference on Software Engineering}}. IEEE,
  \bibinfo{pages}{802--811}.
\newblock
\urldef\tempurl%
\url{https://doi.org/10.1109/ICSE.2013.6606626}
\showDOI{\tempurl}


\bibitem[\protect\citeauthoryear{Kim, Kim, Bissyand{\'e}, Choi, Li, Klein, and
  Le~Traon}{Kim et~al\mbox{.}}{2018}]%
        {kim2018facoy}
\bibfield{author}{\bibinfo{person}{Kisub Kim}, \bibinfo{person}{Dongsun Kim},
  \bibinfo{person}{Tegawend{\'e}~F. Bissyand{\'e}}, \bibinfo{person}{Eunjong
  Choi}, \bibinfo{person}{Li Li}, \bibinfo{person}{Jacques Klein}, {and}
  \bibinfo{person}{Yves Le~Traon}.} \bibinfo{year}{2018}\natexlab{}.
\newblock \showarticletitle{{FaCoY:} a code-to-code search engine}. In
  \bibinfo{booktitle}{\emph{Proceedings of the 40th International Conference on
  Software Engineering}}. ACM, \bibinfo{pages}{946--957}.
\newblock
\urldef\tempurl%
\url{https://doi.org/10.1145/3180155.3180187}
\showDOI{\tempurl}


\bibitem[\protect\citeauthoryear{Koyuncu, Liu, Bissyand{\'e}, Kim, Klein,
  Monperrus, and Traon}{Koyuncu et~al\mbox{.}}{2020}]%
        {koyuncu2020fixminer}
\bibfield{author}{\bibinfo{person}{Anil Koyuncu}, \bibinfo{person}{Kui Liu},
  \bibinfo{person}{Tegawend{\'e}~F. Bissyand{\'e}}, \bibinfo{person}{Dongsun
  Kim}, \bibinfo{person}{Jacques Klein}, \bibinfo{person}{Martin Monperrus},
  {and} \bibinfo{person}{Yves~Le Traon}.} \bibinfo{year}{2020}\natexlab{}.
\newblock \showarticletitle{{FixMiner:} Mining relevant fix patterns for
  automated program repair}.
\newblock \bibinfo{journal}{\emph{Empirical Software Engineering}}
  \bibinfo{volume}{25}, \bibinfo{number}{3} (\bibinfo{year}{2020}),
  \bibinfo{pages}{1980--2024}.
\newblock
\urldef\tempurl%
\url{https://doi.org/10.1007/s10664-019-09780-z}
\showDOI{\tempurl}


\bibitem[\protect\citeauthoryear{Le, Lo, Goues, and Grunske}{Le
  et~al\mbox{.}}{2016b}]%
        {le2016learning}
\bibfield{author}{\bibinfo{person}{Tien{-}Duy~B. Le}, \bibinfo{person}{David
  Lo}, \bibinfo{person}{Claire~Le Goues}, {and} \bibinfo{person}{Lars
  Grunske}.} \bibinfo{year}{2016}\natexlab{b}.
\newblock \showarticletitle{A learning-to-rank based fault localization
  approach using likely invariants}. In \bibinfo{booktitle}{\emph{Proceedings
  of the 25th International Symposium on Software Testing and Analysis}}.
  \bibinfo{publisher}{{ACM}}, \bibinfo{pages}{177--188}.
\newblock
\urldef\tempurl%
\url{https://doi.org/10.1145/2931037.2931049}
\showDOI{\tempurl}


\bibitem[\protect\citeauthoryear{Le, Bao, Lo, Xia, Li, and Pasareanu}{Le
  et~al\mbox{.}}{2019}]%
        {le2019reliability}
\bibfield{author}{\bibinfo{person}{Xuan-Bach~D Le}, \bibinfo{person}{Lingfeng
  Bao}, \bibinfo{person}{David Lo}, \bibinfo{person}{Xin Xia},
  \bibinfo{person}{Shanping Li}, {and} \bibinfo{person}{Corina Pasareanu}.}
  \bibinfo{year}{2019}\natexlab{}.
\newblock \showarticletitle{On reliability of patch correctness assessment}. In
  \bibinfo{booktitle}{\emph{Proceedings of the 41st International Conference on
  Software Engineering}}. IEEE, \bibinfo{pages}{524--535}.
\newblock
\urldef\tempurl%
\url{https://doi.org/10.1109/ICSE.2019.00064}
\showDOI{\tempurl}


\bibitem[\protect\citeauthoryear{Le, Chu, Lo, Le~Goues, and Visser}{Le
  et~al\mbox{.}}{2017}]%
        {le2017s3}
\bibfield{author}{\bibinfo{person}{Xuan-Bach~D Le}, \bibinfo{person}{Duc-Hiep
  Chu}, \bibinfo{person}{David Lo}, \bibinfo{person}{Claire Le~Goues}, {and}
  \bibinfo{person}{Willem Visser}.} \bibinfo{year}{2017}\natexlab{}.
\newblock \showarticletitle{S3: syntax-and semantic-guided repair synthesis via
  programming by examples}. In \bibinfo{booktitle}{\emph{Proceedings of the
  11th Joint Meeting on Foundations of Software Engineering}}. ACM,
  \bibinfo{pages}{593--604}.
\newblock
\urldef\tempurl%
\url{https://doi.org/10.1145/3106237.3106309}
\showDOI{\tempurl}


\bibitem[\protect\citeauthoryear{Le, Lo, and Le~Goues}{Le
  et~al\mbox{.}}{2016a}]%
        {le2016history}
\bibfield{author}{\bibinfo{person}{Xuan Bach~D Le}, \bibinfo{person}{David Lo},
  {and} \bibinfo{person}{Claire Le~Goues}.} \bibinfo{year}{2016}\natexlab{a}.
\newblock \showarticletitle{History driven program repair}. In
  \bibinfo{booktitle}{\emph{Proceedings of the 23rd IEEE International
  Conference on Software Analysis, Evolution, and Reengineering}}. IEEE,
  \bibinfo{pages}{213--224}.
\newblock
\urldef\tempurl%
\url{https://doi.org/10.1109/SANER.2016.76}
\showDOI{\tempurl}


\bibitem[\protect\citeauthoryear{Le, Thung, Lo, and Le~Goues}{Le
  et~al\mbox{.}}{2018}]%
        {le2018overfitting}
\bibfield{author}{\bibinfo{person}{Xuan Bach~D Le}, \bibinfo{person}{Ferdian
  Thung}, \bibinfo{person}{David Lo}, {and} \bibinfo{person}{Claire Le~Goues}.}
  \bibinfo{year}{2018}\natexlab{}.
\newblock \showarticletitle{Overfitting in semantics-based automated program
  repair}.
\newblock \bibinfo{journal}{\emph{Empirical Software Engineering}}
  \bibinfo{volume}{23}, \bibinfo{number}{5} (\bibinfo{year}{2018}),
  \bibinfo{pages}{3007--3033}.
\newblock
\urldef\tempurl%
\url{https://doi.org/10.1007/s10664-017-9577-2}
\showDOI{\tempurl}


\bibitem[\protect\citeauthoryear{Le~Goues, Dewey-Vogt, Forrest, and
  Weimer}{Le~Goues et~al\mbox{.}}{2012a}]%
        {le2012systematic}
\bibfield{author}{\bibinfo{person}{Claire Le~Goues}, \bibinfo{person}{Michael
  Dewey-Vogt}, \bibinfo{person}{Stephanie Forrest}, {and}
  \bibinfo{person}{Westley Weimer}.} \bibinfo{year}{2012}\natexlab{a}.
\newblock \showarticletitle{A systematic study of automated program repair:
  Fixing 55 out of 105 bugs for \$8 each}. In
  \bibinfo{booktitle}{\emph{Proceedings of the 34th International Conference on
  Software Engineering}}. IEEE, \bibinfo{pages}{3--13}.
\newblock
\urldef\tempurl%
\url{https://doi.org/10.1109/ICSE.2012.6227211}
\showDOI{\tempurl}


\bibitem[\protect\citeauthoryear{Le~Goues, Nguyen, Forrest, and
  Weimer}{Le~Goues et~al\mbox{.}}{2012b}]%
        {le2012genprog}
\bibfield{author}{\bibinfo{person}{Claire Le~Goues}, \bibinfo{person}{ThanhVu
  Nguyen}, \bibinfo{person}{Stephanie Forrest}, {and} \bibinfo{person}{Westley
  Weimer}.} \bibinfo{year}{2012}\natexlab{b}.
\newblock \showarticletitle{{GenProg}: A generic method for automatic software
  repair}.
\newblock \bibinfo{journal}{\emph{IEEE Transactions on Software Engineering}}
  \bibinfo{volume}{38}, \bibinfo{number}{1} (\bibinfo{year}{2012}),
  \bibinfo{pages}{54--72}.
\newblock
\urldef\tempurl%
\url{https://doi.org/10.1109/TSE.2011.104}
\showDOI{\tempurl}


\bibitem[\protect\citeauthoryear{Le~Goues, Pradel, and Roychoudhury}{Le~Goues
  et~al\mbox{.}}{2019}]%
        {le2019automated}
\bibfield{author}{\bibinfo{person}{Claire Le~Goues}, \bibinfo{person}{Michael
  Pradel}, {and} \bibinfo{person}{Abhik Roychoudhury}.}
  \bibinfo{year}{2019}\natexlab{}.
\newblock \showarticletitle{Automated Program Repair}.
\newblock \bibinfo{journal}{\emph{Commun. ACM}} \bibinfo{volume}{62},
  \bibinfo{number}{12} (\bibinfo{year}{2019}), \bibinfo{pages}{56--65}.
\newblock
\urldef\tempurl%
\url{https://doi.org/10.1145/3318162}
\showDOI{\tempurl}


\bibitem[\protect\citeauthoryear{Lin, Koppel, Chen, and Solar-Lezama}{Lin
  et~al\mbox{.}}{2017}]%
        {lin2017quixbugs}
\bibfield{author}{\bibinfo{person}{Derrick Lin}, \bibinfo{person}{James
  Koppel}, \bibinfo{person}{Angela Chen}, {and} \bibinfo{person}{Armando
  Solar-Lezama}.} \bibinfo{year}{2017}\natexlab{}.
\newblock \showarticletitle{{QuixBugs:} A multi-lingual program repair
  benchmark set based on the Quixey Challenge}. In
  \bibinfo{booktitle}{\emph{Proceedings Companion of the 2017 ACM SIGPLAN
  International Conference on Systems, Programming, Languages, and
  Applications: Software for Humanity}}. ACM, \bibinfo{pages}{55--56}.
\newblock
\urldef\tempurl%
\url{https://doi.org/10.1145/3135932.3135941}
\showDOI{\tempurl}


\bibitem[\protect\citeauthoryear{Liu, Kim, Bissyand{\'e}, Yoo, and
  Le~Traon}{Liu et~al\mbox{.}}{2018a}]%
        {liu2018mining2}
\bibfield{author}{\bibinfo{person}{Kui Liu}, \bibinfo{person}{Dongsun Kim},
  \bibinfo{person}{Tegawend{\'e}~F Bissyand{\'e}}, \bibinfo{person}{Shin Yoo},
  {and} \bibinfo{person}{Yves Le~Traon}.} \bibinfo{year}{2018}\natexlab{a}.
\newblock \showarticletitle{Mining fix patterns for findbugs violations}.
\newblock \bibinfo{journal}{\emph{IEEE Transactions on Software Engineering}}
  (\bibinfo{year}{2018}).
\newblock
\urldef\tempurl%
\url{https://doi.org/10.1109/TSE.2018.2884955}
\showDOI{\tempurl}


\bibitem[\protect\citeauthoryear{Liu, Koyuncu, Bissyand{\'e}, Kim, Klein, and
  Traon}{Liu et~al\mbox{.}}{2019a}]%
        {liu2019you}
\bibfield{author}{\bibinfo{person}{Kui Liu}, \bibinfo{person}{Anil Koyuncu},
  \bibinfo{person}{Tegawend{\'e}~F Bissyand{\'e}}, \bibinfo{person}{Dongsun
  Kim}, \bibinfo{person}{Jacques Klein}, {and} \bibinfo{person}{Yves~Le
  Traon}.} \bibinfo{year}{2019}\natexlab{a}.
\newblock \showarticletitle{You cannot fix what you cannot find! an
  investigation of fault localization bias in benchmarking automated program
  repair systems}. In \bibinfo{booktitle}{\emph{Proceedings of the 12th IEEE
  International Conference on Software Testing, Verification and Validation}}.
  IEEE, \bibinfo{pages}{102--113}.
\newblock
\urldef\tempurl%
\url{https://doi.org/10.1109/ICST.2019.00020}
\showDOI{\tempurl}


\bibitem[\protect\citeauthoryear{Liu, Koyuncu, Kim, and Bissyand{\'e}}{Liu
  et~al\mbox{.}}{2019b}]%
        {liu2019avatar}
\bibfield{author}{\bibinfo{person}{Kui Liu}, \bibinfo{person}{Anil Koyuncu},
  \bibinfo{person}{Dongsun Kim}, {and} \bibinfo{person}{Tegawend{\'e}~F
  Bissyand{\'e}}.} \bibinfo{year}{2019}\natexlab{b}.
\newblock \showarticletitle{{AVATAR:} Fixing semantic bugs with fix patterns of
  static analysis violations}. In \bibinfo{booktitle}{\emph{Proceedings of the
  26th IEEE International Conference on Software Analysis, Evolution and
  Reengineering}}. IEEE, \bibinfo{pages}{456--467}.
\newblock
\urldef\tempurl%
\url{https://doi.org/10.1109/SANER.2019.8667970}
\showDOI{\tempurl}


\bibitem[\protect\citeauthoryear{Liu, Koyuncu, Kim, and Bissyand{\'e}}{Liu
  et~al\mbox{.}}{2019c}]%
        {liu2019tbar}
\bibfield{author}{\bibinfo{person}{Kui Liu}, \bibinfo{person}{Anil Koyuncu},
  \bibinfo{person}{Dongsun Kim}, {and} \bibinfo{person}{Tegawend{\'e}~F.
  Bissyand{\'e}}.} \bibinfo{year}{2019}\natexlab{c}.
\newblock \showarticletitle{{TBar}: Revisiting Template-based Automated Program
  Repair}. In \bibinfo{booktitle}{\emph{Proceedings of the 28th ACM SIGSOFT
  International Symposium on Software Testing and Analysis}}. ACM,
  \bibinfo{pages}{31--42}.
\newblock
\urldef\tempurl%
\url{https://doi.org/10.1145/3293882.3330577}
\showDOI{\tempurl}


\bibitem[\protect\citeauthoryear{Liu, Koyuncu, Kim, Kim, and Bissyand{\'e}}{Liu
  et~al\mbox{.}}{2018b}]%
        {liu2018lsrepair}
\bibfield{author}{\bibinfo{person}{Kui Liu}, \bibinfo{person}{Anil Koyuncu},
  \bibinfo{person}{Kisub Kim}, \bibinfo{person}{Dongsun Kim}, {and}
  \bibinfo{person}{Tegawend{\'e}~F. Bissyand{\'e}}.}
  \bibinfo{year}{2018}\natexlab{b}.
\newblock \showarticletitle{{LSRepair}: Live search of fix ingredients for
  automated program repair}. In \bibinfo{booktitle}{\emph{Proceedings of the
  25th Asia-Pacific Software Engineering Conference ERA Track}}. {IEEE},
  \bibinfo{pages}{658--662}.
\newblock
\urldef\tempurl%
\url{https://doi.org/10.1109/APSEC.2018.00085}
\showDOI{\tempurl}


\bibitem[\protect\citeauthoryear{Liu and Zhong}{Liu and Zhong}{2018}]%
        {liu2018mining}
\bibfield{author}{\bibinfo{person}{Xuliang Liu} {and} \bibinfo{person}{Hao
  Zhong}.} \bibinfo{year}{2018}\natexlab{}.
\newblock \showarticletitle{Mining stackoverflow for program repair}. In
  \bibinfo{booktitle}{\emph{Proceedings of the 25th IEEE International
  Conference on Software Analysis, Evolution and Reengineering}}. IEEE,
  \bibinfo{pages}{118--129}.
\newblock
\urldef\tempurl%
\url{https://doi.org/10.1109/SANER.2018.8330202}
\showDOI{\tempurl}


\bibitem[\protect\citeauthoryear{Long and Rinard}{Long and Rinard}{2016a}]%
        {long2016analysis}
\bibfield{author}{\bibinfo{person}{Fan Long} {and} \bibinfo{person}{Martin
  Rinard}.} \bibinfo{year}{2016}\natexlab{a}.
\newblock \showarticletitle{An analysis of the search spaces for generate and
  validate patch generation systems}. In \bibinfo{booktitle}{\emph{Proceedings
  of the 38th International Conference on Software Engineering}}. IEEE,
  \bibinfo{pages}{702--713}.
\newblock
\urldef\tempurl%
\url{https://doi.org/10.1145/2884781.2884872}
\showDOI{\tempurl}


\bibitem[\protect\citeauthoryear{Long and Rinard}{Long and Rinard}{2016b}]%
        {long2016automatic}
\bibfield{author}{\bibinfo{person}{Fan Long} {and} \bibinfo{person}{Martin
  Rinard}.} \bibinfo{year}{2016}\natexlab{b}.
\newblock \showarticletitle{Automatic patch generation by learning correct
  code}. In \bibinfo{booktitle}{\emph{Proceedings of the 43rd Annual {ACM}
  {SIGPLAN-SIGACT} Symposium on Principles of Programming Languages}},
  Vol.~\bibinfo{volume}{51}. ACM, \bibinfo{pages}{298--312}.
\newblock
\urldef\tempurl%
\url{https://doi.org/10.1145/2837614.2837617}
\showDOI{\tempurl}


\bibitem[\protect\citeauthoryear{Madeiral, Urli, Maia, and Monperrus}{Madeiral
  et~al\mbox{.}}{2019}]%
        {madeiral2019bears}
\bibfield{author}{\bibinfo{person}{Fernanda Madeiral}, \bibinfo{person}{Simon
  Urli}, \bibinfo{person}{Marcelo Maia}, {and} \bibinfo{person}{Martin
  Monperrus}.} \bibinfo{year}{2019}\natexlab{}.
\newblock \showarticletitle{{BEARS:} An Extensible Java Bug Benchmark for
  Automatic Program Repair Studies}. In \bibinfo{booktitle}{\emph{Proceedings
  of the 26th International Conference on Software Analysis, Evolution and
  Reengineering}}. IEEE, \bibinfo{pages}{468--478}.
\newblock
\urldef\tempurl%
\url{https://doi.org/10.1109/SANER.2019.8667991}
\showDOI{\tempurl}


\bibitem[\protect\citeauthoryear{Mann and Whitney}{Mann and Whitney}{1947}]%
        {mann1947test}
\bibfield{author}{\bibinfo{person}{Henry~B Mann} {and}
  \bibinfo{person}{Donald~R. Whitney}.} \bibinfo{year}{1947}\natexlab{}.
\newblock \showarticletitle{On a {{Test}} of {{Whether}} One of {{Two Random
  Variables}} Is {{Stochastically Larger}} than the {{Other}}}.
\newblock \bibinfo{journal}{\emph{The Annals of Mathematical Statistics}}
  \bibinfo{volume}{18}, \bibinfo{number}{1} (\bibinfo{year}{1947}),
  \bibinfo{pages}{50--60}.
\newblock
\urldef\tempurl%
\url{https://doi.org/10.1214/aoms/1177730491}
\showDOI{\tempurl}


\bibitem[\protect\citeauthoryear{Martinez and Monperrus}{Martinez and
  Monperrus}{2016}]%
        {martinez2016astor}
\bibfield{author}{\bibinfo{person}{Matias Martinez} {and}
  \bibinfo{person}{Martin Monperrus}.} \bibinfo{year}{2016}\natexlab{}.
\newblock \showarticletitle{{ASTOR:} a program repair library for Java (demo)}.
  In \bibinfo{booktitle}{\emph{Proceedings of the 25th International Symposium
  on Software Testing and Analysis}}. ACM, \bibinfo{pages}{441--444}.
\newblock
\urldef\tempurl%
\url{https://doi.org/10.1145/2931037.2948705}
\showDOI{\tempurl}


\bibitem[\protect\citeauthoryear{Martinez and Monperrus}{Martinez and
  Monperrus}{2018}]%
        {martinez2018ultra}
\bibfield{author}{\bibinfo{person}{Matias Martinez} {and}
  \bibinfo{person}{Martin Monperrus}.} \bibinfo{year}{2018}\natexlab{}.
\newblock \showarticletitle{Ultra-Large Repair Search Space with Automatically
  Mined Templates: the Cardumen Mode of Astor}. In
  \bibinfo{booktitle}{\emph{Proceedings of the 10th International Symposium on
  Search Based Software Engineering}}. Springer, \bibinfo{pages}{65--86}.
\newblock
\urldef\tempurl%
\url{https://doi.org/10.1007/978-3-319-99241-9\_3}
\showDOI{\tempurl}


\bibitem[\protect\citeauthoryear{Monperrus}{Monperrus}{2014}]%
        {monperrus2014critical}
\bibfield{author}{\bibinfo{person}{Martin Monperrus}.}
  \bibinfo{year}{2014}\natexlab{}.
\newblock \showarticletitle{A critical review of automatic patch generation
  learned from human-written patches: essay on the problem statement and the
  evaluation of automatic software repair}. In
  \bibinfo{booktitle}{\emph{Proceedings of the 36th International Conference on
  Software Engineering}}. ACM, \bibinfo{pages}{234--242}.
\newblock
\urldef\tempurl%
\url{https://doi.org/10.1145/2568225.2568324}
\showDOI{\tempurl}


\bibitem[\protect\citeauthoryear{Monperrus}{Monperrus}{2018a}]%
        {monperrus2018automatic}
\bibfield{author}{\bibinfo{person}{Martin Monperrus}.}
  \bibinfo{year}{2018}\natexlab{a}.
\newblock \showarticletitle{Automatic software repair: {A} bibliography}.
\newblock \bibinfo{journal}{\emph{Comput. Surveys}} \bibinfo{volume}{51},
  \bibinfo{number}{1} (\bibinfo{year}{2018}), \bibinfo{pages}{17:1--17:24}.
\newblock
\urldef\tempurl%
\url{https://doi.org/10.1145/3105906}
\showDOI{\tempurl}


\bibitem[\protect\citeauthoryear{Monperrus}{Monperrus}{2018b}]%
        {monperrus2018living}
\bibfield{author}{\bibinfo{person}{Martin Monperrus}.}
  \bibinfo{year}{2018}\natexlab{b}.
\newblock \showarticletitle{The living review on automated program repair}. In
  \bibinfo{booktitle}{\emph{HAL/archives-ouvertes. fr, Technical Report}}.
\newblock


\bibitem[\protect\citeauthoryear{Motwani, Sankaranarayanan, Just, and
  Brun}{Motwani et~al\mbox{.}}{2018}]%
        {motwani2018automated}
\bibfield{author}{\bibinfo{person}{Manish Motwani}, \bibinfo{person}{Sandhya
  Sankaranarayanan}, \bibinfo{person}{Ren{\'e} Just}, {and}
  \bibinfo{person}{Yuriy Brun}.} \bibinfo{year}{2018}\natexlab{}.
\newblock \showarticletitle{Do automated program repair techniques repair hard
  and important bugs?}
\newblock \bibinfo{journal}{\emph{Empirical Software Engineering}}
  \bibinfo{volume}{23}, \bibinfo{number}{5} (\bibinfo{year}{2018}),
  \bibinfo{pages}{2901--2947}.
\newblock
\urldef\tempurl%
\url{https://doi.org/10.1007/s10664-017-9550-0}
\showDOI{\tempurl}


\bibitem[\protect\citeauthoryear{Nguyen, Qi, Roychoudhury, and Chandra}{Nguyen
  et~al\mbox{.}}{2013}]%
        {nguyen2013semfix}
\bibfield{author}{\bibinfo{person}{Hoang Duong~Thien Nguyen},
  \bibinfo{person}{Dawei Qi}, \bibinfo{person}{Abhik Roychoudhury}, {and}
  \bibinfo{person}{Satish Chandra}.} \bibinfo{year}{2013}\natexlab{}.
\newblock \showarticletitle{{SemFix}: Program repair via semantic analysis}. In
  \bibinfo{booktitle}{\emph{Proceedings of the 35th International Conference on
  Software Engineering}}. IEEE, \bibinfo{pages}{772--781}.
\newblock
\urldef\tempurl%
\url{https://doi.org/10.1109/ICSE.2013.6606623}
\showDOI{\tempurl}


\bibitem[\protect\citeauthoryear{Pearson, Campos, Just, Fraser, Abreu, Ernst,
  Pang, and Keller}{Pearson et~al\mbox{.}}{2017}]%
        {pearson2017evaluating}
\bibfield{author}{\bibinfo{person}{Spencer Pearson}, \bibinfo{person}{Jos{\'e}
  Campos}, \bibinfo{person}{Ren{\'e} Just}, \bibinfo{person}{Gordon Fraser},
  \bibinfo{person}{Rui Abreu}, \bibinfo{person}{Michael~D. Ernst},
  \bibinfo{person}{Deric Pang}, {and} \bibinfo{person}{Benjamin Keller}.}
  \bibinfo{year}{2017}\natexlab{}.
\newblock \showarticletitle{Evaluating and improving fault localization}. In
  \bibinfo{booktitle}{\emph{Proceedings of the 39th International Conference on
  Software Engineering}}. \bibinfo{publisher}{{ACM}},
  \bibinfo{pages}{609--620}.
\newblock
\urldef\tempurl%
\url{https://doi.org/10.1109/ICSE.2017.62}
\showDOI{\tempurl}


\bibitem[\protect\citeauthoryear{Qi, Mao, Lei, Dai, and Wang}{Qi
  et~al\mbox{.}}{2014}]%
        {qi2014strength}
\bibfield{author}{\bibinfo{person}{Yuhua Qi}, \bibinfo{person}{Xiaoguang Mao},
  \bibinfo{person}{Yan Lei}, \bibinfo{person}{Ziying Dai}, {and}
  \bibinfo{person}{Chengsong Wang}.} \bibinfo{year}{2014}\natexlab{}.
\newblock \showarticletitle{The strength of random search on automated program
  repair}. In \bibinfo{booktitle}{\emph{Proceedings of the 36th International
  Conference on Software Engineering}}. ACM, \bibinfo{pages}{254--265}.
\newblock
\urldef\tempurl%
\url{https://doi.org/10.1145/2568225.2568254}
\showDOI{\tempurl}


\bibitem[\protect\citeauthoryear{Qi, Mao, Lei, and Wang}{Qi
  et~al\mbox{.}}{2013}]%
        {qi2013using}
\bibfield{author}{\bibinfo{person}{Yuhua Qi}, \bibinfo{person}{Xiaoguang Mao},
  \bibinfo{person}{Yan Lei}, {and} \bibinfo{person}{Chengsong Wang}.}
  \bibinfo{year}{2013}\natexlab{}.
\newblock \showarticletitle{Using automated program repair for evaluating the
  effectiveness of fault localization techniques}. In
  \bibinfo{booktitle}{\emph{Proceedings of the 22nd International Symposium on
  Software Testing and Analysis}}. ACM, \bibinfo{pages}{191--201}.
\newblock
\urldef\tempurl%
\url{https://doi.org/10.1145/2483760.2483785}
\showDOI{\tempurl}


\bibitem[\protect\citeauthoryear{Qi, Long, Achour, and Rinard}{Qi
  et~al\mbox{.}}{2015}]%
        {qi2015analysis}
\bibfield{author}{\bibinfo{person}{Zichao Qi}, \bibinfo{person}{Fan Long},
  \bibinfo{person}{Sara Achour}, {and} \bibinfo{person}{Martin Rinard}.}
  \bibinfo{year}{2015}\natexlab{}.
\newblock \showarticletitle{An analysis of patch plausibility and correctness
  for generate-and-validate patch generation systems}. In
  \bibinfo{booktitle}{\emph{Proceedings of the 24th International Symposium on
  Software Testing and Analysis}}. ACM, \bibinfo{pages}{24--36}.
\newblock
\urldef\tempurl%
\url{https://doi.org/10.1145/2771783.2771791}
\showDOI{\tempurl}


\bibitem[\protect\citeauthoryear{Saha, Lyu, Lam, Yoshida, and Prasad}{Saha
  et~al\mbox{.}}{2018}]%
        {saha2018bugs}
\bibfield{author}{\bibinfo{person}{Ripon Saha}, \bibinfo{person}{Yingjun Lyu},
  \bibinfo{person}{Wing Lam}, \bibinfo{person}{Hiroaki Yoshida}, {and}
  \bibinfo{person}{Mukul Prasad}.} \bibinfo{year}{2018}\natexlab{}.
\newblock \showarticletitle{Bugs.jar: A large-scale, diverse dataset of
  real-world java bugs}. In \bibinfo{booktitle}{\emph{Proceedings of the 15th
  IEEE/ACM International Conference on Mining Software Repositories}}. ACM,
  \bibinfo{pages}{10--13}.
\newblock
\urldef\tempurl%
\url{https://doi.org/10.1145/3196398.3196473}
\showDOI{\tempurl}


\bibitem[\protect\citeauthoryear{Saha, Lyu, Yoshida, and Prasad}{Saha
  et~al\mbox{.}}{2017}]%
        {saha2017elixir}
\bibfield{author}{\bibinfo{person}{Ripon~K Saha}, \bibinfo{person}{Yingjun
  Lyu}, \bibinfo{person}{Hiroaki Yoshida}, {and} \bibinfo{person}{Mukul~R
  Prasad}.} \bibinfo{year}{2017}\natexlab{}.
\newblock \showarticletitle{{ELIXIR:} Effective object-oriented program
  repair}. In \bibinfo{booktitle}{\emph{Proceedings of the 32nd IEEE/ACM
  International Conference on Automated Software Engineering}}. IEEE,
  \bibinfo{pages}{648--659}.
\newblock
\urldef\tempurl%
\url{https://doi.org/10.1109/ASE.2017.8115675}
\showDOI{\tempurl}


\bibitem[\protect\citeauthoryear{Saha, Saha, and Prasad}{Saha
  et~al\mbox{.}}{2019}]%
        {saha2019harnessing}
\bibfield{author}{\bibinfo{person}{Seemanta Saha}, \bibinfo{person}{Ripon~K
  Saha}, {and} \bibinfo{person}{Mukul~R Prasad}.}
  \bibinfo{year}{2019}\natexlab{}.
\newblock \showarticletitle{Harnessing evolution for multi-hunk program
  repair}. In \bibinfo{booktitle}{\emph{Proceedings of the 41st International
  Conference on Software Engineering}}. IEEE, \bibinfo{pages}{13--24}.
\newblock
\urldef\tempurl%
\url{https://doi.org/10.1109/ICSE.2019.00020}
\showDOI{\tempurl}


\bibitem[\protect\citeauthoryear{Smith, Barr, Le~Goues, and Brun}{Smith
  et~al\mbox{.}}{2015}]%
        {smith2015cure}
\bibfield{author}{\bibinfo{person}{Edward~K Smith}, \bibinfo{person}{Earl~T
  Barr}, \bibinfo{person}{Claire Le~Goues}, {and} \bibinfo{person}{Yuriy
  Brun}.} \bibinfo{year}{2015}\natexlab{}.
\newblock \showarticletitle{Is the cure worse than the disease? overfitting in
  automated program repair}. In \bibinfo{booktitle}{\emph{Proceedings of the
  10th Joint Meeting on Foundations of Software Engineering}}. ACM,
  \bibinfo{pages}{532--543}.
\newblock
\urldef\tempurl%
\url{https://doi.org/10.1145/2786805.2786825}
\showDOI{\tempurl}


\bibitem[\protect\citeauthoryear{Sobreira, Durieux, Madeiral, Monperrus, and
  de~Almeida~Maia}{Sobreira et~al\mbox{.}}{2018}]%
        {sobreira2018dissection}
\bibfield{author}{\bibinfo{person}{Victor Sobreira}, \bibinfo{person}{Thomas
  Durieux}, \bibinfo{person}{Fernanda Madeiral}, \bibinfo{person}{Martin
  Monperrus}, {and} \bibinfo{person}{Marcelo de Almeida~Maia}.}
  \bibinfo{year}{2018}\natexlab{}.
\newblock \showarticletitle{Dissection of a bug dataset: Anatomy of 395 patches
  from Defects4J}. In \bibinfo{booktitle}{\emph{Proceedings of the 25th
  International Conference on Software Analysis, Evolution and Reengineering}}.
  IEEE, \bibinfo{pages}{130--140}.
\newblock
\urldef\tempurl%
\url{https://doi.org/10.1109/SANER.2018.8330203}
\showDOI{\tempurl}


\bibitem[\protect\citeauthoryear{Wang, Wen, Chen, Yi, and Mao}{Wang
  et~al\mbox{.}}{2019a}]%
        {wang2019different}
\bibfield{author}{\bibinfo{person}{Shangwen Wang}, \bibinfo{person}{Ming Wen},
  \bibinfo{person}{Liqian Chen}, \bibinfo{person}{Xin Yi}, {and}
  \bibinfo{person}{Xiaoguang Mao}.} \bibinfo{year}{2019}\natexlab{a}.
\newblock \showarticletitle{How Different Is It Between Machine-Generated and
  Developer-Provided Patches? An Empirical Study on The Correct Patches
  Generated by Automated Program Repair Techniques}. In
  \bibinfo{booktitle}{\emph{Proceedings of the 13th International Symposium on
  Empirical Software Engineering and Measurement}}.
  \bibinfo{publisher}{{IEEE}}, \bibinfo{pages}{1--12}.
\newblock
\urldef\tempurl%
\url{https://doi.org/10.1109/ESEM.2019.8870172}
\showDOI{\tempurl}


\bibitem[\protect\citeauthoryear{Wang, Wen, Mao, and Yang}{Wang
  et~al\mbox{.}}{2019b}]%
        {wang2019attention}
\bibfield{author}{\bibinfo{person}{Shangwen Wang}, \bibinfo{person}{Ming Wen},
  \bibinfo{person}{Xiaoguang Mao}, {and} \bibinfo{person}{Deheng Yang}.}
  \bibinfo{year}{2019}\natexlab{b}.
\newblock \showarticletitle{Attention please: Consider Mockito when evaluating
  newly proposed automated program repair techniques}. In
  \bibinfo{booktitle}{\emph{Proceedings of the 23rd Evaluation and Assessment
  on Software Engineering}}. ACM, \bibinfo{pages}{260--266}.
\newblock
\urldef\tempurl%
\url{https://doi.org/10.1145/3319008.3319349}
\showDOI{\tempurl}


\bibitem[\protect\citeauthoryear{Weimer, Nguyen, Le~Goues, and Forrest}{Weimer
  et~al\mbox{.}}{2009}]%
        {weimer2009automatically}
\bibfield{author}{\bibinfo{person}{Westley Weimer}, \bibinfo{person}{ThanhVu
  Nguyen}, \bibinfo{person}{Claire Le~Goues}, {and} \bibinfo{person}{Stephanie
  Forrest}.} \bibinfo{year}{2009}\natexlab{}.
\newblock \showarticletitle{Automatically finding patches using genetic
  programming}. In \bibinfo{booktitle}{\emph{Proceedings of the 31st
  International Conference on Software Engineering}}. IEEE,
  \bibinfo{pages}{364--374}.
\newblock
\urldef\tempurl%
\url{https://doi.org/10.1109/ICSE.2009.5070536}
\showDOI{\tempurl}


\bibitem[\protect\citeauthoryear{Wen, Chen, Wu, Hao, and Cheung}{Wen
  et~al\mbox{.}}{2018}]%
        {wen2018context}
\bibfield{author}{\bibinfo{person}{Ming Wen}, \bibinfo{person}{Junjie Chen},
  \bibinfo{person}{Rongxin Wu}, \bibinfo{person}{Dan Hao}, {and}
  \bibinfo{person}{Shing-Chi Cheung}.} \bibinfo{year}{2018}\natexlab{}.
\newblock \showarticletitle{Context-aware patch generation for better automated
  program repair}. In \bibinfo{booktitle}{\emph{Proceedings of the 40th
  International Conference on Software Engineering}}. ACM,
  \bibinfo{pages}{1--11}.
\newblock
\urldef\tempurl%
\url{https://doi.org/10.1145/3180155.3180233}
\showDOI{\tempurl}


\bibitem[\protect\citeauthoryear{Wen, Wu, Liu, Tian, Xie, Cheung, and Su}{Wen
  et~al\mbox{.}}{2019}]%
        {wen2019exploring}
\bibfield{author}{\bibinfo{person}{Ming Wen}, \bibinfo{person}{Rongxin Wu},
  \bibinfo{person}{Yepang Liu}, \bibinfo{person}{Yongqiang Tian},
  \bibinfo{person}{Xuan Xie}, \bibinfo{person}{Shing-Chi Cheung}, {and}
  \bibinfo{person}{Zhendong Su}.} \bibinfo{year}{2019}\natexlab{}.
\newblock \showarticletitle{Exploring and {{Exploiting}} the {{Correlations
  Between Bug}}-Inducing and {{Bug}}-Fixing {{Commits}}}. In
  \bibinfo{booktitle}{\emph{Proceedings of the 27th {{ACM Joint Meeting}} on
  {{European Software Engineering Conference}} and {{Symposium}} on the
  {{Foundations}} of {{Software Engineering}}}}. \bibinfo{publisher}{{ACM}},
  \bibinfo{pages}{326--337}.
\newblock
\urldef\tempurl%
\url{https://doi.org/10.1145/3338906.3338962}
\showDOI{\tempurl}


\bibitem[\protect\citeauthoryear{White, Tufano, Martinez, Monperrus, and
  Poshyvanyk}{White et~al\mbox{.}}{2019}]%
        {white2019sorting}
\bibfield{author}{\bibinfo{person}{Martin White}, \bibinfo{person}{Michele
  Tufano}, \bibinfo{person}{Matias Martinez}, \bibinfo{person}{Martin
  Monperrus}, {and} \bibinfo{person}{Denys Poshyvanyk}.}
  \bibinfo{year}{2019}\natexlab{}.
\newblock \showarticletitle{Sorting and transforming program repair ingredients
  via deep learning code similarities}. In
  \bibinfo{booktitle}{\emph{Proceedings of the IEEE 26th International
  Conference on Software Analysis, Evolution and Reengineering}}. IEEE,
  \bibinfo{pages}{479--490}.
\newblock
\urldef\tempurl%
\url{https://doi.org/10.1109/SANER.2019.8668043}
\showDOI{\tempurl}


\bibitem[\protect\citeauthoryear{Wilcoxon}{Wilcoxon}{1945}]%
        {wilcoxon1945individual}
\bibfield{author}{\bibinfo{person}{F. Wilcoxon}.}
  \bibinfo{year}{1945}\natexlab{}.
\newblock \showarticletitle{Individual Comparisons by Ranking Methods}.
\newblock \bibinfo{journal}{\emph{Biometrics Bulletin}} \bibinfo{volume}{1},
  \bibinfo{number}{6} (\bibinfo{year}{1945}), \bibinfo{pages}{80--83}.
\newblock


\bibitem[\protect\citeauthoryear{Xin and Reiss}{Xin and Reiss}{2017}]%
        {xin2017leveraging}
\bibfield{author}{\bibinfo{person}{Qi Xin} {and} \bibinfo{person}{Steven~P
  Reiss}.} \bibinfo{year}{2017}\natexlab{}.
\newblock \showarticletitle{Leveraging syntax-related code for automated
  program repair}. In \bibinfo{booktitle}{\emph{Proceedings of the 32nd
  IEEE/ACM International Conference on Automated Software Engineering}}. IEEE,
  \bibinfo{pages}{660--670}.
\newblock
\urldef\tempurl%
\url{https://doi.org/10.1109/ASE.2017.8115676}
\showDOI{\tempurl}


\bibitem[\protect\citeauthoryear{Xiong, Liu, Zeng, Zhang, and Huang}{Xiong
  et~al\mbox{.}}{2018}]%
        {xiong2018identifying}
\bibfield{author}{\bibinfo{person}{Yingfei Xiong}, \bibinfo{person}{Xinyuan
  Liu}, \bibinfo{person}{Muhan Zeng}, \bibinfo{person}{Lu Zhang}, {and}
  \bibinfo{person}{Gang Huang}.} \bibinfo{year}{2018}\natexlab{}.
\newblock \showarticletitle{Identifying patch correctness in test-based program
  repair}. In \bibinfo{booktitle}{\emph{Proceedings of the 40th International
  Conference on Software Engineering}}. ACM, \bibinfo{pages}{789--799}.
\newblock
\urldef\tempurl%
\url{https://doi.org/10.1145/3183519.3183540}
\showDOI{\tempurl}


\bibitem[\protect\citeauthoryear{Xiong, Wang, Yan, Zhang, Han, Huang, and
  Zhang}{Xiong et~al\mbox{.}}{2017}]%
        {xiong2017precise}
\bibfield{author}{\bibinfo{person}{Yingfei Xiong}, \bibinfo{person}{Jie Wang},
  \bibinfo{person}{Runfa Yan}, \bibinfo{person}{Jiachen Zhang},
  \bibinfo{person}{Shi Han}, \bibinfo{person}{Gang Huang}, {and}
  \bibinfo{person}{Lu Zhang}.} \bibinfo{year}{2017}\natexlab{}.
\newblock \showarticletitle{Precise condition synthesis for program repair}. In
  \bibinfo{booktitle}{\emph{Proceedings of the 39th IEEE/ACM International
  Conference on Software Engineering}}. IEEE, \bibinfo{pages}{416--426}.
\newblock
\urldef\tempurl%
\url{https://doi.org/10.1109/ICSE.2017.45}
\showDOI{\tempurl}


\bibitem[\protect\citeauthoryear{Xuan, Martinez, Demarco, Clement, Marcote,
  Durieux, Le~Berre, and Monperrus}{Xuan et~al\mbox{.}}{2017}]%
        {xuan2017nopol}
\bibfield{author}{\bibinfo{person}{Jifeng Xuan}, \bibinfo{person}{Matias
  Martinez}, \bibinfo{person}{Favio Demarco}, \bibinfo{person}{Maxime Clement},
  \bibinfo{person}{Sebastian~Lamelas Marcote}, \bibinfo{person}{Thomas
  Durieux}, \bibinfo{person}{Daniel Le~Berre}, {and} \bibinfo{person}{Martin
  Monperrus}.} \bibinfo{year}{2017}\natexlab{}.
\newblock \showarticletitle{Nopol: Automatic repair of conditional statement
  bugs in java programs}.
\newblock \bibinfo{journal}{\emph{IEEE Transactions on Software Engineering}}
  \bibinfo{volume}{43}, \bibinfo{number}{1} (\bibinfo{year}{2017}),
  \bibinfo{pages}{34--55}.
\newblock
\urldef\tempurl%
\url{https://doi.org/10.1109/TSE.2016.2560811}
\showDOI{\tempurl}


\bibitem[\protect\citeauthoryear{Yang, Zhikhartsev, Liu, and Tan}{Yang
  et~al\mbox{.}}{2017}]%
        {yang2017better}
\bibfield{author}{\bibinfo{person}{Jinqiu Yang}, \bibinfo{person}{Alexey
  Zhikhartsev}, \bibinfo{person}{Yuefei Liu}, {and} \bibinfo{person}{Lin Tan}.}
  \bibinfo{year}{2017}\natexlab{}.
\newblock \showarticletitle{Better test cases for better automated program
  repair}. In \bibinfo{booktitle}{\emph{Proceedings of the 11th Joint Meeting
  on Foundations of Software Engineering}}. ACM, \bibinfo{pages}{831--841}.
\newblock
\urldef\tempurl%
\url{https://doi.org/10.1145/3106237.3106274}
\showDOI{\tempurl}


\bibitem[\protect\citeauthoryear{Yi, Tan, Mechtaev, B{\"o}hme, and
  Roychoudhury}{Yi et~al\mbox{.}}{2018}]%
        {yi2018correlation}
\bibfield{author}{\bibinfo{person}{Jooyong Yi}, \bibinfo{person}{Shin~Hwei
  Tan}, \bibinfo{person}{Sergey Mechtaev}, \bibinfo{person}{Marcel B{\"o}hme},
  {and} \bibinfo{person}{Abhik Roychoudhury}.} \bibinfo{year}{2018}\natexlab{}.
\newblock \showarticletitle{A correlation study between automated program
  repair and test-suite metrics}.
\newblock \bibinfo{journal}{\emph{Empirical Software Engineering}}
  \bibinfo{volume}{23}, \bibinfo{number}{5} (\bibinfo{year}{2018}),
  \bibinfo{pages}{2948--2979}.
\newblock
\urldef\tempurl%
\url{https://doi.org/10.1007/s10664-017-9552-y}
\showDOI{\tempurl}


\bibitem[\protect\citeauthoryear{Yuan and Banzhaf}{Yuan and Banzhaf}{2018}]%
        {yuan2018arja}
\bibfield{author}{\bibinfo{person}{Yuan Yuan} {and} \bibinfo{person}{Wolfgang
  Banzhaf}.} \bibinfo{year}{2018}\natexlab{}.
\newblock \showarticletitle{{ARJA:} Automated Repair of Java Programs via
  Multi-Objective Genetic Programming}.
\newblock \bibinfo{journal}{\emph{IEEE Transactions on Software Engineering}}
  (\bibinfo{year}{2018}).
\newblock
\urldef\tempurl%
\url{https://doi.org/10.1109/TSE.2018.2874648}
\showDOI{\tempurl}


\end{thebibliography}

\end{document}